\documentclass[aip,pop,sd,reprint,10pt,amsmath,amssymb,graphicx]{revtex4-1}

\usepackage[capitalize]{cleveref}
\usepackage[T1]{fontenc}
\usepackage{float}
\usepackage{rotating}
\usepackage[normalem]{ulem}
\usepackage{amsmath}
\usepackage{textcomp}
\usepackage{color}
\usepackage{marvosym}
\usepackage{wasysym}
\usepackage{amssymb}
\usepackage{array}

\usepackage{bm}
\usepackage{bigints}

\newcommand{\beq}{\begin{equation}}
\newcommand{\eeq}{\end{equation}}
\newcommand{\bfig}{\begin{figure}[htbp]}
	\newcommand{\efig}{\end{figure}}

\newcommand{\ben}{\begin{eqnarray}}
\newcommand{\een}{\end{eqnarray}}
\usepackage{graphicx}
\usepackage{color}
\usepackage{subfig}
\usepackage{setspace}

\usepackage{epsfig}
\usepackage[titletoc,toc,title]{appendix}
\usepackage{tikz}
\usetikzlibrary{positioning}
\usepackage{mathtools}
\usepackage{environ}
\NewEnviron{myequation}{%
	\begin{equation}
	\scalebox{1.2}{$\BODY$}
	\end{equation}
}

\newtagform{Alph}[]()
\newtagform{roman}[]()
\newtagform{scroman}[][]

\begin{document}
	\title{ Prediction of temperature barriers in weakly collisional plasmas by a Lagrangian Coherent Structures computational tool}
	\author{G. Di Giannatale}\affiliation{Consorzio RFX (CNR, ENEA, INFN, Università di Padova, Acciaierie Venete SpA), Corso Stati Uniti 4, Padova, Italy}
	\author{D. Bonfiglio}\affiliation{Consorzio RFX (CNR, ENEA, INFN, Università di Padova, Acciaierie Venete SpA), Corso Stati Uniti 4, Padova, Italy}
	\author{S. Cappello}\affiliation{Consorzio RFX (CNR, ENEA, INFN, Università di Padova, Acciaierie Venete SpA), Corso Stati Uniti 4, Padova, Italy}
	\author{L. Chac\'on}\affiliation{Los Alamos National Laboratory, Los Alamos, New Mexico 87545, USA}
	\author{M. Veranda}\affiliation{Consorzio RFX (CNR, ENEA, INFN, Università di Padova, Acciaierie Venete SpA), Corso Stati Uniti 4, Padova, Italy}

	\bigskip
	
\begin{abstract}
Analysis of Lagrangian Coherent Structures (LCSs) has been showed to be a valid mathematical approach to explain the formation of transport barriers  in magnetized plasmas. Such LCSs, borrowed from fluid dynamics theory, can be considered as the hidden skeleton of the system and can be used for studying a wide spectrum of transport mechanisms {even} in plasmas. In this paper, we demonstrate that such structures can be particularly useful for underlying the hidden paths governing the motion of magnetic field lines in chaotic magnetic fields. \\
To perform such an analysis, we developed a numerical tool able to detect LCSs for {a general dynamical system}. The tool is able to deal with general coordinate systems and it is shown to match with other techniques already used to analyse chaotic magnetic fields, {e.g.} connection length. {After the description of the computational tool,} we focus on the heat transport equation and the comparison between  temperature profile and {topology of the} LCSs. We provide evidence that numerical simulations are able to reproduce the temperature profiles similar to those observed in RFP experiments and that our tool successfully predicts the location of temperature gradients.
The results suggest that, inside the chaotic region, the field-lines motion is far from stochastic and that the presence of hidden patterns allows the development of high temperature gradients. 

\end{abstract}

\maketitle

\section{Introduction}

The understanding of transport phenomena is notoriously one of the most challenging tasks in the physics of magnetized plasmas. 
Depending on the device, different physical processes appear to dominate transport. {Tokamaks generally 
feature micro-turbulence related transport. Moreover, sometimes tokamaks suffer from  Magnetohydrodynamics (MHD) instabilities
leading to loss of confinement and in some cases to disruptions}. Not optimized Stellarators suffer from large neoclassical transport. Reversed
field pinch devices (RFP) are characterized by MHD activity leading to  chaotic magnetic
fields and {related chaotic transport}. 
Among these transport mechanisms (microturbulence, neoclassical, MHD-induced chaotic), historically the chaotic one has been deemed the worst. To some extent, all magnetic confinement concepts feature {chaotic transport.}  For example, during sawtooth oscillations \cite{sawtooth,Waddell_1976} 
magnetic chaos develops\cite{Igochine_2007,Igochine_2008} in the core of tokamaks and chaotic transport takes place.  Chaotic transport in tokamaks is also thought to be present during the development of Edge Localized Modes (ELM) and the magnetic edge oscillations recently observed during the Quasi-H mode discharges (QH-regime) in the DIII-D experiment \cite{Garofalo2015,Liu_2015}. {Moreover, for ITER-relevant scenarios, {subdominant} linearly stable microtearing modes were identified {as a possible}  mechanism for the development of magnetic stochasticity and transport in gyrokinetic simulations of electromagnetic ion temperature gradient (ITG) driven plasma microturbulence \cite{Hatch2013}. This stochasticity disrupts zonal flows associated with ITG turbulence saturation, which increases the level of the turbulence itself\cite{Terry2013}. Finally, magnetic chaos develops  during disruptions\cite{Boozer_2012} and when magnetic perturbations are applied for ELM mitigation and control\cite{Kim_2020}. 

{Stellarators have a  magnetic field that is intrinsically 3D and the devices are prone to develop edge chaotic regions that increase with plasma pressure.}

In fusion plasmas, heat transport is dominated by electrons, which move
essentially along magnetic field lines. Thus, it is inherently anisotropic with the parallel thermal conductivity $\chi_\parallel$ considerably larger than the  perpendicular one $\chi_\perp$: theory and experiments\cite{braginski_anisotropia,H_lzl_2009} suggest $\chi_\parallel/\chi_\perp$ exceeds $10^{10}$.
For this reason, the study of the topology and structure of the chaotic magnetic field lines acquires a crucial importance.

Even though magnetic chaos may lead to a degradation of the confinement properties,
magnetic barriers can emerge also in a chaotic region, {preventing ergodicity}
\cite{Misguich_2002,digiannatale1,digiannatale2,borgogno2011barriers,rubino2015detection,Pegoraro_2019,Veranda_2017,Veranda_2019}. Such barriers may account for the formation of {internal} electron {transport} barriers that have been experimentally observed in the RFX device\cite{Lorenzini}. Similar results can be found even in tokamaks and stellarators: after applying a resonant magnetic perturbation\cite{Kim_2020}, a pedestal still remains even though a chaotic magnetic region formed in the edge{\cite{Volpe_2012}}. The same is found in QH-regimes where the edge naturally  develops  a chaotic region\cite{Garofalo2015}; a chaotic boundary due to the 3D plasma response features temperature gradients in stellarators\cite{Suzuki2013} {too.}

{The divergence-free nature of the magnetic field allows to cast the equations of the field lines trajectories in the form of a hamiltonian dynamical system\cite{boozer1981plasma,cary1983noncanonical}}, as it will be briefly described later in this paper. This  allows  exploiting  the techniques used for the study of Hamiltonian and more general dynamical systems. In fact, often chaotic magnetic fields are studied using the Poincar\'e plot technique. Unfortunately, when the system features extended chaotic seas, the Poincar\'e plot could hide preferential transport channels and  partial barriers that survive in the chaotic regions. 

In order to  analyse  chaotic transport phenomena we use the concept of Lagrangian Coherent Structures (LCSs), introduced by G. Haller in
the context of transport processes in fluid flows \cite{haller2011variational,haller2015lagrangian,farazmand2014shearless}. LCSs, which are a finite time generalization of the manifolds of the systems, are structures ruling the behaviour of the dynamical system. Indeed, over the finite time span which characterizes the LCSs, such structures govern the motion of the system and act as the hidden skeleton of the dynamics, governing where and how {field lines (for our studies)} can move. 
In addition, LCSs locate boundaries of regions inside which
fast mixing phenomena take place. 
The goal of this paper is twofold: 
\begin{enumerate}
	\item presenting a numerical tool able to compute LCSs for  2D non autonomous (2.5D) systems, i.e., dynamical systems governed by the equation:
	\begin{equation}\label{general_system}
	{\bf \dot{x}} = {\bf v} ({\bf x},t) \qquad {\bf x}=(x,y) \text{.}
	\end{equation}
	
	\item  showing that LCSs correspond to gradients of connection length and temperature in 3D nonlinear MHD {realistic} numerical simulations, demonstrating that, even in a chaotic magnetic field, temperature gradients may form, as it is observed in experiments\cite{Sakakibara_2008,Martines_2011,Lorenzini,cappello_2008}.
	
\end{enumerate}

Systems like the one  in Eq. \ref{general_system} appear very often in {fusion} plasma physics and  the LCSs technique {may be applied} in other areas such as  micro-turbulence \cite{Padberg_2007}, fast particles transport, beam plasma instability\cite{carlevaro_2015,carlevaro_montani_falessi_2020} (where the system is aperiodic), reduced Vlasov equations.

The paper is divided in two parts.
In the first part, we {present} the  algorithm to detect the LCS. Such part is devoted to the reader who is interested to the methods and techniques we adopt. Particular attention has been devoted to the description of mathematical objects governing the deformation of points of the {dynamical} system (magnetic field lines in  our case) for an arbitrary geometry. \\
In the second part, we show the effectiveness of this tool {for realistic} magnetic fields obtained from 3D MHD numerical simulations. In particular, we show that LCSs are a useful technique to locate regions where transport of field lines is reduced, and that the LCS govern the transport of such field lines. Finally, we will show how  {such } LCSs affect  the temperature transport in a weakly  collisional and anisotropic plasma: LCSs predict the location of temperature gradients. {Our results} counter  the conventional wisdom that high temperature gradients cannot be sustained inside a chaotic region. Similar arguments have been exposed exploiting {\itshape ghost surfaces}\cite{hudson_ghost}.\\
The paper is structured as follows. Section II {briefly recall}  LCSs   mathematical definition. The third section {presents the LCS tool}. {Here we describe three essential implements for LCS detection: computation of Cauchy-Green strain tensor, auxiliary grid and backward integration}. The fourth section illustrates the results of the tool. In particular, LCSs are compared with connection length (a relation with the Cantor set will be {also} discussed) and with the temperature {map obtained by a heat equation solver\cite{t3d}  that computes the temperature for a given, {fixed} magnetic field.} We {also} discuss  the temperature {map obtained} for a chaotic RFP magnetic field. The last section contains the conclusions of the work.

\section{Lagrangian Coherent Structures as maximal repulsion-attraction material lines}
We begin by recalling the definition of LCSs  by Haller \cite{haller2011variational}, which  considers the LCSs as the most attracting or repelling material surfaces (curves in 2D non autonomous systems). 
Following the approach used in \onlinecite{falessi2015lagrangian}, we consider a dynamical system in 2D phase space ${\boldsymbol  x} = (x,y)$ with a continuous differentiable flow map 
\begin{equation}\label{map}
{\boldsymbol \Phi}_{t_0}^t({\boldsymbol  x}_0)={\boldsymbol  x}(t,t_0,{\boldsymbol  x}_0).
\end{equation}
Two neighbouring points ${\boldsymbol x}_0$ and ${\boldsymbol x}_0+ {\boldsymbol   \delta   \boldsymbol x}_0$ evolve into ${\boldsymbol x}$ and ${\boldsymbol x}+ {\boldsymbol \delta \boldsymbol x}$ according to
\begin{equation}
{\boldsymbol  \delta \boldsymbol x} ={\bf F}_{t_0}^t \,  {\boldsymbol\delta  \boldsymbol x}_0 + \mathcal{O}(||{\boldsymbol\delta  \boldsymbol x}_0 ||^2).
\end{equation}
Here, ${\bf F}$ is the deformation gradient of the transformation, $ {{\bf F}_{t_0}^t} = \nabla \, \Phi^t_{t_0}$. Consider a curve $\gamma_0$ and at  each point  ${\boldsymbol  x}_0 \in \gamma_0$ define the unit tangent and normal  vectors ${\boldsymbol  e}_0$ and ${\boldsymbol  n}_0$. In the  interval $[t_0,t]$, $\gamma_0$ evolves into $\gamma_t$, and  ${\boldsymbol  x}_0\in \gamma_0$ into ${\boldsymbol  x}_t \in \gamma_t$. 
The tangent vector ${\boldsymbol  e}_0$ evolves, by means of the linearised dynamics, into
\begin{equation}
{\boldsymbol  e}_t = \frac{{{\bf F}^t_{t_0}({\boldsymbol  x}_0)\, {\boldsymbol  e}_0}}{{ [ {\boldsymbol  e}_0\,  {\boldsymbol  C}^t_{t_0}({\boldsymbol  x}_0) \, {\boldsymbol  e}_0]^{1/2}}},
\end{equation}
where ${\bf  C}^t_{t_0}({\boldsymbol  x}_0)\equiv {\bf F}^T\, {\bf F}$  is the \emph{Cauchy-Green strain tensor (superscript $T$ stands for transposed) } which  describes the deformation into an ellipse, as a first order approximation, of an arbitrarily small circle {(blob)} of initial conditions (i.c.) centred at ${\boldsymbol  x}_0$. 
Regarding ${\bf n}$, in general the mapping does not preserve the angle between vectors and therefore usually ${\bf n}_t $ differs from ${\bf F}^t_{t_0} {\bf n}_0$. However, using the orthogonality condition between ${\bf n}_0, {\bf e}_0$ and ${\bf n}_t, {\bf e}_t$ it is possible to obtain:
\begin{equation}
{\boldsymbol  n}_t = \frac{  {\left( {\bf F}^{t_0}_{t}\right)^T{\boldsymbol  n}_0}}{{ [ {\boldsymbol  n}_0\,  {\boldsymbol  C}^{-1}({\boldsymbol  x}_0) \, {\boldsymbol  n}_0]^{1/2}}} \text{,}
\end{equation}
where   ${\bf  C}^{-1} ({\boldsymbol  x}_0 ) = {\bf  C}^{t_0}_t ({\boldsymbol  x}_0 ) $ and the time interval subscripts have been suppressed, as will be the case in the following formulae. 
\,The \emph{repulsion rate} $\rho^t_{t_0}({\boldsymbol  x}_0,{\boldsymbol  n}_0)$ of a curve $\gamma_0$ is defined (see  Ref.\onlinecite{haller2011variational,falessi2015lagrangian}) as the {rate} at which  points initially   near  ${\boldsymbol  x}_0 \in \gamma_0$   increase  their distance from the curve \textbf{$[t_0,t]$}:   
\begin{alignat}{2}\label{rep_ratio}
\rho^t_{t_0}({\boldsymbol  x}_0,{\boldsymbol  n}_0)&={\boldsymbol  n}_t \, {\bf F}^t_{t_0}({\boldsymbol  x}_0) \, {\boldsymbol  n}_0
=\\[.2cm]
&=[ {\boldsymbol  n}_0\,  {\bf  C}^{-1}({\boldsymbol  x}_0) \, {\boldsymbol  n}_0]^{-1/2}=  [ {\boldsymbol  n}_t\,  {\bf  C}({\boldsymbol  x}_0) \, {\boldsymbol  n}_t]^{1/2} \, . \nonumber 
\end{alignat}

An LCS over a finite time interval $\left[t_0, t_0+T \right]$ is defined as a {\itshape material}  line {(i.e., a codimension-one
invariant surface in the extended phase space of a dynamical
system)} along which the repulsion rate is pointwise maximal. This leads, as shown in  Refs. \onlinecite{haller2011variational,falessi2015lagrangian}, to the following definitions.\\
Being $\lambda_{max}, \lambda_{min}$ the eigenvalues of ${\bf C}$, {with  ${\boldsymbol  \xi}_{max} , {\boldsymbol  \xi}_{min} $ the corresponding eigenvectors}, a {\itshape material} line (or surface for higher dimensionality systems)  is called  a repulsive Weak Lagrangian Coherent Structure (WLCS) if it  satisfies the conditions:
\begin{alignat}{2}
&\text{(I)}& \qquad \lambda_{min}<\lambda_{max},\quad \lambda_{max}>1  \qquad,\label{condizionelambda}  \\[.2cm]
&\text{(II)}& \qquad {\boldsymbol e}_0=  {\boldsymbol \xi}_{min}  \qquad   \qquad  \qquad ,  \label{condizione2} \\[.2cm]
 &\text{(III)}&\qquad  \quad   \quad   \, {\boldsymbol  \xi}_{max} \cdot {\boldsymbol  \nabla} \lambda_{max} =0 \qquad  \qquad . 
\label{numericallyaproblem}
\end{alignat}

A WLCS that satisfies   at each point the additional condition
\begin{equation}\label{maxrep}
 \text{(IV)} \qquad  \; {\boldsymbol  \xi}_{max} \, \cdot  {\boldsymbol   \nabla }^2 \lambda_{max}\,  \cdot \, {\boldsymbol   \xi}_{max}  <0 \quad \; \; 
\end{equation}
is called a {repulsive} Lagrangian Coherent Structure. {Attractive LCSs are defined as repulsive LCSs of the backward-time dynamics}. In reference to the blob of i.c. that is deformed into an ellipse, {the above conditions read as follows}: (I)  there is at least one direction of stretching;   (II)  repelling LCSs are  strain lines of the {\bf C} tensor; (III)  the gradient of the largest eigenvalue is along the curve;  (IV) $\lambda_{max}$ decreases perpendicularly to the curve. 


\section{Description of the LCS computational tool }\label{section3}
This section presents the  numerical tool {developed} to detect LCSs for the systems governed by  dynamics described by Eq. \ref{general_system}.\\
The first subsection, {${\bf A}$}, describes the main steps of the algorithm. Then, we dedicate subsections $\bf{B,C,D}$ for describing three aspects:  {\bf B}) the way the geometry is taken into account; {\bf C}) the way adopting an auxiliary grid helps  to {\itshape numerically} improve the calculations of the quantities related to the CG tensor; {\bf D}) how  {we} improve the LCSs computation by back-integrating in time those curves satisfying  conditions (I-IV).  An additional prescription concerning the way  condition III is checked is explained in the {\bf Appendix  \ref{appendiceA}}.
\subsection{General description}\label{general_description}

An overview of the key steps to compute LCS can be seen in Table \ref{table_1}.\\
After  defining a grid of initial conditions (i.c.),  the trajectories of the i.c.  (${\bf x}_i(0)$ with $i=0,\dots, N$) are computed  solving  equation \ref{general_system}. {How dense is the grid of i.c., that is the choice of N, decides the quality of LCS (in the section \ref{connection_length_section}, we give details about how we choose $N$).} 
 Such trajectories are used to compute the gradient, ${\bf F} = \nabla \Phi$,  of the map $\Phi$ associated to the system. When this matrix is obtained, the Cauchy-Green tensor, $ {\bf C} = {\bf F}^T {\bf F} $, is computed with its eigenvalues ($\lambda_{min}, \lambda_{max}$) and eigenvectors ($\boldsymbol{\xi}_{min}, \boldsymbol{\xi}_{max}$) (Sec. \ref{sezione_def_grad} is devoted to such computations in general geometries).\\
{Now that we have computed $\lambda_{i}$ and $\boldsymbol{\xi}_{i}$ in the whole grid, we seek the curves satisfying  conditions (I-IV)}. Condition (II) says that the tangent vector to an LCS is parallel to $\boldsymbol{\xi}_{min}$. We can enforce such a condition by construction by solving the equation: 
\begin{equation}\label{condizione2_compu}
	\frac{\text{d}\gamma(s)}{\text{d}s} = \boldsymbol{\xi}_{min} \qquad \gamma(0) = {\bf x}_0
\end{equation}
where $s$ is the arclength of the LCS and $\gamma$ is a curve in the plane (x,y).  After the computation of several $\gamma(s)_i$, we check whether  conditions I, III and IV are satisfied along the curves. The curves $\gamma_i$ satisfying them can be considered as LCS.

In the following  we describe how we choose the starting points for the integration of eqn \ref{condizione2_compu} and then we describe a numerical problem that usually occurs during its integration.\\ 

\begin{table*}
	\fontsize{9}{11}\selectfont
	\begin{tabular}{|p{5mm}|l|l|}
		\hline
		&$\qquad \qquad \qquad \qquad \qquad \qquad \qquad \qquad \qquad \qquad$ TASK
		& in depth description\\
		\hline
		1. & Define a vector field ${\bf v}({\bf x},t)$ for the ODE ${\bf \dot{x}} = {\bf v} ({\bf x},t)$
		& \\
		\hline
		2. & Decide the number of initial conditions (points of the main grid) (and add the auxiliary grid) &  Sec III A (C)\\
		\hline
		3. & Compute the finite time trajectories ${\bf x}_i(t)$ of the initial conditions & \\
		\hline
		4. & Compute Cauchy-Green strain tensor with eigenvalues and eigenvectors, respectively  $\lambda_{i}$, $ {\boldsymbol \xi}_i \qquad$ & Sec \ref{sezione_def_grad}\\
		\hline
		5. & Set the initial conditions (i.c.) $\gamma(s=0)$ for the integration $\frac{\text{d}\gamma(s)}{\text{d}s} = \boldsymbol{\xi}_{min}$  & Sec \ref{general_description}-a\\
		\hline
		6. & Integrate the equation $\frac{\text{d}\gamma(s)}{\text{d}s} = \boldsymbol{\xi}_{min}$ for each i.c.  & Sec \ref{general_description}-b\\
		\hline
		7. & Check  conditions III and IV, i.e. ${\boldsymbol  \xi}_{max} \cdot {\boldsymbol  \nabla} \lambda_{max} =0$ and  ${\boldsymbol  \xi}_{max} \, \cdot  {\boldsymbol   \nabla }^2 \lambda_{max}\,  \cdot \, {\boldsymbol   \xi}_{max}  <0$   & Appendix \ref{appendiceA} \\
		\hline
		8. & Estimate a hierarchy for the LCS according to eqn. \ref{mean_repulsion} & Sec \ref{general_description}-a\\
		\hline	
		9. & Backward integrate the LCS  (for periodic systems only) & Sec  \ref{back_int}\\
		\hline	
	\end{tabular}

	\caption{Overview of sequence of computations to detect LCSs. On the right the section where the task is deeply analysed is shown.}
	\label{table_1}
\end{table*}

\paragraph{Choice of the "starting point" and LCS hierarchy} \label{choosing_point}
In order to integrate {  equation \ref{condizione2_compu}}, we choose to start from local maxima of $\lambda_{max} $, since according to eqns. \ref{numericallyaproblem},\ref{maxrep}, the LCS has to pass through maxima of $\lambda_{max}$. 
Unfortunately, a chaotic field has many $\lambda_{max}$ maxima: by integrating the above condition for each local maximum we would find too many structures to be practical. For these reasons, we restrict our attention to absolute maxima of $\lambda_{max} $  within a {chosen area-neighbourhood} $\mathcal{A}$ (we use $\mathcal{A} \sim 1/200 $ of the whole domain,  $A_{tot}$). If two maxima are found inside the area $\mathcal{A}$, we disregard the point with the lower value.  This approach is only used to {reduce to a workable}  number the initial conditions for the integration of eq. \ref{condizione2_compu}. However, $\mathcal{A} \sim 1/200 \,A_{tot}$  {still gives so many structures ($\sim 200$) that may} confuse
the physical information which we wish to extract. {To obtain the most significant information, a strategy is needed}. 

{To do so, a first order  approach adopted in Ref. \onlinecite{digiannatale2} is increasing the neighbourhood $\mathcal{A}$. In this way fewer maxima $\lambda_{max}$ (only the strongest ones) are taken as initial condition for eqn \ref{condizione2_compu}.} This procedure tends to promote LCSs going through a strong local value of $\lambda_{max} $ but does not assure that, being $\gamma_0$ the LCS, for the whole structure   $\lambda_{max}(\gamma_0) $ is large. There is no guarantee that such LCSs are  the most significant for the dynamics.


{Here the approach is different. {We still use an initial finite $\mathcal{A}$ for the choice of starting point (integrating for all $\lambda_{max}$ maxima would be impractical)}, but to extract the most significant LCS we adopt an LCSs hierarchy  built according to the repulsion rate of the LCSs.}
The repulsion rate along an LCS $\gamma$ is defined, according to eqn \ref{rep_ratio} in the case ${\bf n}_0 = {\boldsymbol{\xi}}_{max}$, as:
\begin{equation}\label{mean_repulsion}
R(\gamma) = \frac{ \bigintss_{\gamma}{\sqrt{\lambda_{max}} \; ds}}{\bigintss_{\gamma} ds} \,.
\end{equation}
LCSs featuring the  highest values of repulsion rate are considered as the most important ones.

%

\paragraph{Condition II: LCS curve integration} \label{curve_integration}
 
{According to eqn \ref{condizione2_compu}}, the curve develops following the eigenvectors ${\boldsymbol  \xi}_{min} $. The integration ends either when the curve reaches a pre-defined maximum length, when it reaches the domain edge or when the algorithm detects that the curves is moving following a random walk meaning lack of convergence, as we will describe in the next paragraph. After the integration ends, it is repeated from the same starting point but with the phases-inverted eigenvectors. The curve integration is performed, owing to the vector transformation explained in following section (Sec. \ref{sezione_def_grad}), on {a} logical grid. In this way, the grid is Cartesian and the distance between points in vertical and horizontal direction is equal to 1 {(corresponding to lengths of the cells of the logical grid)}.  A point of the grid corresponds to the index of a certain initial condition. \\
A problem that is often encountered during the integration is that  new curve positions behave as a random process. This can happen when eigenvectors are not accurate and often it happens in those points that spread too fast and where the computation of the CG tensor is very sensitive. In a chaotic field, we cannot get rid of this problem: exact mathematical equations for trajectories do not exist and the  computation is thus subjected to an intrinsic numerical error. 
{Such errors may lead to nearby points of the grid having eigenvectors with random phase, causing a random walk during the curve integration.}
To overcome such problem, after the CG computation and eigenvalues-eigenvectors extraction, we check  the {\itshape quality} of computation. {Being  $\mathbb{T}$ a volume stretch factor threshold}, if $1/\mathbb{T} < \lambda_{1} \lambda_2 < \mathbb{T}$ then we accept the eigenvector at that point for the interpolation; otherwise, it is rejected.
The reader should be aware that this criterion can be used only in  volume preserving maps; in which the determinant of ${\bf C}$ is 1. In other kind of systems, other mathematical-physical constraints can be taken into account to assess the consistency of the computation.

{The algorithm described here computes the LCSs according the conditions I-IV. It goes beyond previous studies  focusing on equations I-II \cite{digiannatale2} or finding points with maxima value of $\lambda_{max}$\cite{rubino2015detection,borgogno2011barriers,Veranda_2017}}.
Computing LCSs using only equation \ref{condizione2_compu} may lead to  false  LCSs. This is shown in appendix A. Similarly, identifying LCS as points with maximum values of $\lambda_{max}$ may lead to misleading results, as shown in Ref \onlinecite{haller2011variational}. The algorithm we use here solves these issues.

\subsection{Deformation gradient and CG calculation}\label{sezione_def_grad}
Using coordinates $\Theta^i$ for the reference position (before deformation), {the current (after deformation) position can be written as} $u^i = \Phi^i(\Theta^1,\Theta^2,\Theta^3,t)$.  {Note that} the coordinate system at current and reference position can be  different. \\
The deformation gradient ${\bf F}({\bf X},t) = \nabla  \Phi({\bf X},t)$  is defined as the gradient of the map giving the motion of a {point} ${\bf X}$
occupying the position ${\bf x}$ at time t, where ${\bf X, x}$ are respectively the initial and final position in Cartesian coordinates. In curvilinear coordinates, this takes the general form \cite{St_Hartmann,Marsden}: 
\begin{equation}\label{gradient_deform}
{\bf F}({\bf X},t) = \frac{\partial \Phi^i}{\partial \Theta^k} \: {\bf e}_{u_i} \: {\bf e}^{\Theta_k} \,.
\end{equation}

The tensor in eq.  \ref{gradient_deform} is expressed using normalized  basis vectors $\hat{{\bf e}}_{u_i}, \; \hat{{\bf e}}^{\Theta_i}$. Thus,  the gradient is expressed as: 
\begin{equation}
{\bf F} = F^i_{\cdot j} \, \hat{{\bf e}}_{u_i} \; \hat{{\bf e}}^{\Theta_k} \qquad \text{with } \qquad F^i_{\cdot j} = \frac{\partial \Phi^i}{\partial \Theta^k} \;  \sqrt{g_{ii} \, G^{kk}} \, ,
\end{equation}
where $g_{ii} = {\bf e}_{u_i } \cdot {\bf e}_{u_i }$ and $G^{ii} = {\bf e}^{\Theta_i } \cdot {\bf e}^{\Theta_i }$.
Note that ${\bf e}_{u_i },\,{\bf e}^{\Theta_k}  $ vectors do not refer to the same basis, with
 ${\bf e}_u$ referring to the final state and ${\bf e}^\Theta$ to the initial one. This may appear strange, but it respects completely the role of ${\bf F}$
 that maps an initial distance $\delta {\bf x}_0$ (defined on the initial metrics) to the final distance after the motion $\delta {\bf x}_1$  (expressed using the final metrics).
The deformation gradient is used to compute the CG tensor, ${\bf C } = {\bf F}^T {\bf F}$, which  gives information about contraction and dilatation of distances during the {evolution} under the map.\\
In case of non-orthonormal metric (that is $\bf{e}_{u_i} \cdot {\bf e}_{u_j} \not= \bf{\delta}_{i,j}$) special attention has to be given to the computation. \\
Indicating ${\bf F} = F^i_{\cdot \; k} \,\hat{{\bf e}}_{u_i }\, \hat{{\bf e}}^{\Theta_k}$ then ${\bf F}^T = (F^T)^{\cdot\; i}_{k} \,\hat{{\bf e}}^{\Theta_k}\, \hat{{\bf e}}_{u_i}$  where  $ \left(F^T \right)^{\cdot\; i}_k = F ^{i}_{\cdot \; k}$.\\
Now the CG tensor can be written as:
\begin{alignat}{2} \label{CG_good}
{\bf C} &= {\bf F}^T {\bf F} = \left(F^T \right)^{\cdot\; i}_{k} \; \hat{{\bf e}}^{\Theta_{k}} \hat{{\bf e}}_{u_i} \; F ^{p}_{\cdot\; q} \; \hat{{\bf e}}_{u_p} \hat{{\bf e}}^{\Theta_{q}} =\\[.2cm]
&= \left( F ^{i}_{\cdot\;  k}  F ^{p}_{\cdot\; q} \; \hat{g}_{i,p} \right) \;  \hat{{\bf e}}^{\Theta_{k}} \hat{{\bf e}}^{\Theta_{q}} = C_{k,q} \; \hat{{\bf e}}^{\Theta_{k}} \hat{{\bf e}}^{\Theta_{q}} \, , \nonumber
\end{alignat}
where $ \hat{g}_{i,p}  $ is the normalized metric tensor 	$\hat{g}_{i,p} = \hat{{\bf e}}_{u_i} \cdot \hat{ {\bf e}}_{u_p}$. The reader should note that  the basis vectors of the CG tensor are those referring to the
initial state.\\
Now to compute the eigenvectors it is necessary to change the basis vectors of the CG tensor. In fact by  definition the eigenvectors are those vectors $\xi$ satisfying ${\bf C} \;\boldsymbol{\xi} = \lambda \; \boldsymbol{\xi} $ and thus the form ${\bf C} = C_{k,q} \; \hat{{\bf e}}^{\Theta_{k}} \hat{{\bf e}}^{\Theta_{q}}$ cannot be used since $C_{i,j}\, \hat{{\bf e}}^{\Theta_i} \,\hat{{\bf e}}^{\Theta_j}  \; \xi^{k} \hat{{\bf e}}_{\Theta_k} = C_{i,k} \, \xi^k \, \hat{{\bf e}}^{\Theta_i}  \not = \lambda \;\xi^i \hat{{\bf e}}_{\Theta_i}$ (the basis is different and standard  matrix techniques cannot be used!).\\
To overcome the problem the CG tensor is transformed into a mixed tensor, and it is expressed as ${\bf C} = C^i_{\cdot \; k} \hat{{\bf e}}_{\Theta_i}\, \hat{{\bf e}}^{\Theta_k}$. In this way:
\begin{equation}
C^i_{\cdot \; k} \hat{{\bf e}}_{\Theta_i}\, \hat{{\bf e}}^{\Theta_k} \; \xi^j \hat{{\bf e}}_{\Theta_j} = \lambda \; \xi^{i} \hat{{\bf e}}_{\Theta_i} \, .
\end{equation}\label{eigenvectors}
At this point we can compute eigenvalues and eigenvectors. Then, since the integration of LCSs is done on a logical grid, we need to perform some operations to trace the eigenvectors in the logical-space. \\
Given the general arc-length: $\text{d}{\bf R} = \sqrt{g_{ii}} \text{d}u^i \hat{{\bf e}}_i$ the eigenvectors are mapped upon the logical grid according to:
\begin{equation} 
\xi^i_{norm} = \frac{\xi^i}{\sqrt{g_{ii}} \; \Delta u^i } \, ,
\end{equation}	
where $\Delta u^i $ is the distance between two points of the main grid on the $ u^i $ axis. Now it does not matter what the geometry is and the integration is performed on a {\itshape logical}-Cartesian grid with unity mesh spacing. In this way, during the integration, the curve moves inside squares defined by indices in a Cartesian logical grid. The eigenvectors are linearly interpolated using the four closest surrounding points.  \\
How the several quantities read in cylindrical coordinate system, which will be used in the cases described in Sec.\ref{risultati}, it is described in Appendix \ref{App_B}.

With this formalism, we overcome previous studies\cite{rubino2015detection,Pegoraro_2019} on similar topics in cylindrical geometry where the authors  do not take into account the right formula of the  deformation tensor (i.e., eqn \ref{deformation_gradient_cyl}). 
However, when the field line radial displacements are small (i.e., chaos not fully developed)  the lack of such formalism for ${\bf F}$ should not lead to substantial changes, and  the results that such works present can be still considered a good estimate.
\subsection{Auxiliary grid}\label{auxiliary_grid_section}
Using an auxiliary grid allows computing  the ${\bf F} $ tensor much more accurately, and accordingly ${\bf C}$ and all the important quantities  related to it, namely the $\lambda_{i}, {\boldsymbol{\xi}_i}$. For each point of the main grid (corresponding to the initial positions of trajectories used to compute the LCSs), there are four auxiliary points disposed as in Fig.\ref{auxiliary_grid}. 
The points belonging to the auxiliary grid are evolved under the flow map and their final position is only used to compute the deformation gradient and then the CG tensor in the points of the main grid.

\begin{figure}[h!]
	\centering
	\includegraphics[height=4cm,width=7cm]{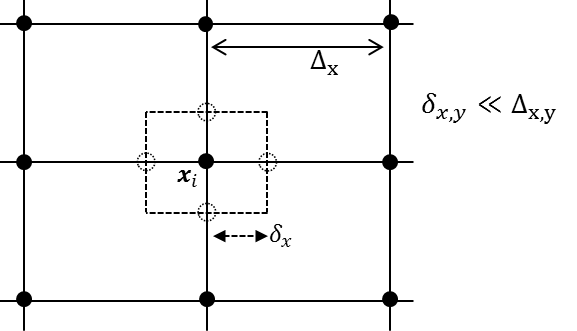}
	\caption{Illustration of the main grid (filled circles) and the auxiliary grid (empty circles) used in the computation of the derivative of the flow map.  The point ${\bf x}_i$ is a point of the main grid and the surrounding  points indicated with empty circles are the points of the auxiliary grids (of the point ${\bf x}_i$).}
	\label{auxiliary_grid}
\end{figure}

Beyond the ${\bf C}$ computation itself, using the auxiliary grid is  in line with the LCS theory:  constraints \ref{condizionelambda}-\ref{maxrep} are obtained by considering a linear evolution of the tangent and normal vectors ${\bf e,n}$ {and} inasmuch as the displacement remains small, the linearised dynamics holds.

Being $N_{ag}^x$ and $N_{ag}^y$ the number of  points of the main grid along x and y-direction, the total number of points that are evolved under the flow map $\Phi$ is $ N_{tot} = N_{ag}^x \, N_{ag}^y \, 5$. But the eigenvectors are computed only in $N_{ag}^x \, N_{ag}^y $ points. Thus, if the same total number of i.c  $ N_{tot}$ is evolved under the flow map $\Phi$, the use of the auxiliary grid reduces to 1/5 the number of points where eigenvectors are available.
Computationally speaking, the step size in the numerical integration of eqn \ref{integration_grid} increases. The consequence could be a deviation from the "real" trajectory that may grow during the integration. However, numerical evidence\cite{tesi_phd_digiannatale} shows that it is more convenient to have fewer points but with a more accurate eigenvector computation than to have more points with inaccurate  eigenvectors. \\
In order to illustrate the accuracy improvements that the auxiliary grid may bring, we take the  standard map \cite{chirikov1,chirikov2} as an example.  The standard map is the volume preserving discrete dynamical system defined by: 
\begin{equation}\label{standard_map}
p' = p - \frac{k}{2 \pi} \sin(2 \pi x) \; , \qquad x' = x + p' \quad \text{(mod \, 1)} \;.
\end{equation}
In these equations, $x$ and $p$ are the two degrees of freedom ($x', p'$ the evolution at the following map iteration) and $k$ is the "chaos parameter": increasing k increases the chaos. In figure \ref{poincare_k_096} the Poincar\'e plot  is shown for $k = 0.96$.
\begin{figure}
	\begin{tikzpicture}[trim left=-4.5cm]
	\node[node distance=1cm] (russell) at (0,0)
	{\includegraphics[height=5.cm,width=8.5cm]{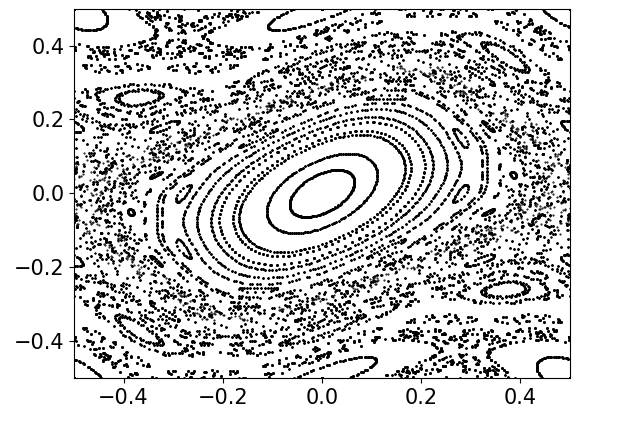}};
	\node[below of=russell, node distance=2.4cm, anchor=center,xshift=2mm] {\Large{$x$}};
	\node[left of=russell, node distance=4.2cm, anchor=center,yshift=2mm] {\Large {$p$}};
	\end{tikzpicture}
	\caption{Poincar\'e plot of the standard map. $k = 0.96$.}
	\label{poincare_k_096}
\end{figure}
Being volume preserving,  $\det {\bf F} = 1$ and so is for $\det {\bf C}$. In order to check whether  the computation is properly carried out, one can check that the eigenvalues product is equal to 1: $\lambda_{max} \, \lambda_{min} = 1$.
For a system with an explicit map (like the standard map) the computation of $\bf{F}$ (and $\bf{C}$) can be done analytically, but in order to test the tool all the computations are done numerically using trajectories.
Of course, for a chaotic system where the derivatives are computed numerically, it is impossible to satisfy this exact requirement, so in general we check $\mathbb{T}^{-1} < \lambda_{max} \, \lambda_{min} < \mathbb{T}$, {where $\mathbb{T}$  is a volume stretch factor threshold}. The reader should be aware that such a condition is much more difficult to satisfy when the  map is not known and the trajectory integration has to be performed numerically (like in a field-line tracing).  
Fig. \ref{auxiliary_grid_vp} shows the huge improvement of such a computation when the auxiliary grid is used. \\
In the left panel we can see that a lot of points break the area-preserving condition. Using the auxiliary grid (right panel), most of the points become area preserving and this ensures a much better computation of the eigenvectors and consequently of the LCSs. 

\begin{figure}
	\begin{tikzpicture}[trim left=-4.5cm]
	\node[node distance=1cm] (russell) at (0,0)
	{\includegraphics[height=4.cm,width=9.5cm]{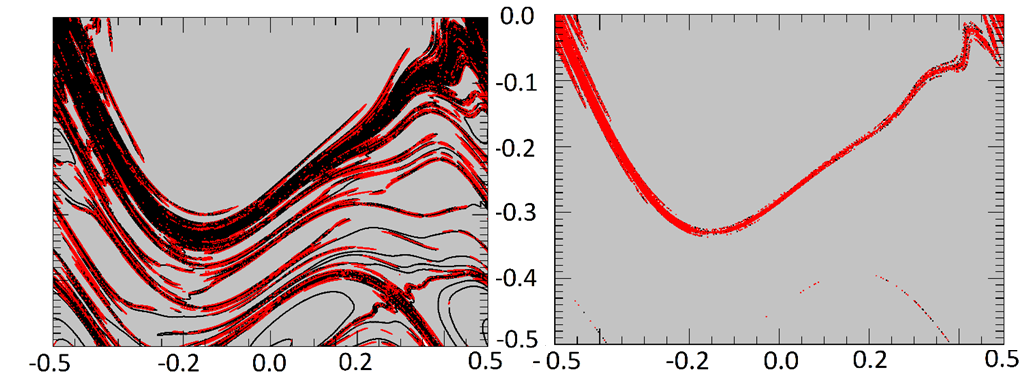}};
	\node[below of=russell, node distance=2.4cm, anchor=center] {\Large{$x$}};
	\node[left of=russell, node distance=4.6cm, anchor=center] {\Large {$p$}};
	\end{tikzpicture}
	\caption{Points in which the numerical volume preserving condition is not satisfied
		for a finite time of 20 map iterations for $k = 0.96$. In red points where  $\lambda_{max}\, \lambda_{min} < 1/3$, in black points where  $\lambda_{max}\, \lambda_{min} > 3$, in gray  volume preserving points. Left panel: without grid, $Nx = 2000, Np = 2000$. Right panel: with auxiliary grid, $N_x = 1000, N_y =1000, \delta_{x,y} = 10^{-12}$.}
	\label{auxiliary_grid_vp}
\end{figure}

\subsection{Backward integration}\label{back_int}
One of the {delicate aspects} in the LCS computation concerns the fact that, due to the finite time interval {(over which LCS are computed)} and due to numerical errors, it is not possible to draw the full {\itshape real}-LCS shape. By {\itshape real}-LCS we mean the real mathematical object governing the dynamics of the system. In order to solve this problem, when the curves satisfying the conditions (I-IV) have been obtained, they are evolved backward, that is the trajectories of the points belonging to the LCSs are evolved under the flow map $\Phi$ reversing the sign of ``time'' in Eq. \ref{map} (for our studies,  the direction along the toroidal direction during the field line trajectories computation\cite{digiannatale1}). 
To explain properly such concept,  let us to refer to a Hamiltonian system, where it is known that LCSs try to mark stable manifolds, or at least their equivalent for finite time. 
When the system features a big island, the LCSs  resemble the  stable manifold part until the few primary homoclinic points \cite{rom1990transport} ($\mu_n$, with usually $n=1,2$) depending on the shape of the manifold, on the finite time used and on the level of chaos (stronger chaos leads to degradation of numerical computation).\\
In order to have structures approaching as much as possible the real LCSs, we use, when it is possible, the backward evolution of the structures. Let us  take an LCS $\gamma(s)$. The backward evolution of such curve under a finite negative time $-t$ leads to a new curve $\sigma(s) = \Phi^{-\hat{t}} \left( \gamma(s) \right)$. If the dynamical system is periodic, such curve $\sigma(s)$ is still an LCS for the same dynamical system $\Phi^t_{t_0}$. This is true since $\gamma(s)$ has the property: $\Phi^t(\gamma) \to \tilde{\bf{x}}$ as $t  \to \infty$, where $\tilde{\bf{x}}$ is an X-point of the system. Then, calling $t_1 = t + \hat{t}$
\begin{equation}
\Phi^t_{t_1}(\sigma) \to \tilde{\bf{x}}    \qquad {\text as} \qquad t_1 \to \infty \quad .
\end{equation}
The resulting curve $\sigma$ has two improvements with respect to $\gamma$: it is longer and more intricate with respect to $\gamma$ (so it better describes the dynamical system) and it reduces the difference with the real LCS.\\
Noting $d(\cdot,\cdot)$ the distance between points of the dynamical system, the first improvement is related to the fact that repelling LCSs experiment, for forward dynamics, an exponential shrinking $\text{d}(\Phi^t({\bf x_1}),\Phi^t({\bf x_2})) \propto \text{e}^{-\alpha t}$, and thus they experiment an exponential stretching for the backward dynamics. Dealing with a map, such shrinking (stretching for backward dynamics)  can be written as  $\text{d}(\Phi^{t+1}({\bf x_1}),\Phi^{t+1}({\bf x_2})) < L \; \text{d}(\Phi^{t}({\bf x_1}),\Phi^{t}({\bf x_2}))$.  \\
Concerning to the second improvement, it comes from the fact that repelling LCSs behave as attracting LCSs for the backward dynamics \cite{haller2011variational,haller2015lagrangian} and thus the error between the computed LCS and the real one decreases.\\
{In figure \ref{backward_int}, the effect of the backward integration for the standard map is shown.  In red, we highlight an LCS obtained satisfying all the LCS conditions above, and in black the backward evolution of the red curve. It is evident how such a LCS acquires a more intricate shape when the backward integration is applied. In particular, it is possible to see that, after the backward integration, the curve is more suitable to describe the complexity of the chaotic transport. Accordingly, backward-integrating the LCS get closer and closer to the real stable manifolds, which are the real invariant curves governing the infinite-time dynamics.\\
In this plot, we focused on the resonances of higher order where the islands have very small amplitude: such a procedure is able to catch even the lobes at very low spatial scales. Of course this happens as long as the integrator for the trajectories is accurate. For the standard map, the integrator can be considered exact because the map governing the dynamics is known (eqn \ref{standard_map}). }
\begin{figure}
	\begin{tikzpicture}[trim left=-4.5cm]
	\node[node distance=0.5cm] (russell) at (0,0)
	{\includegraphics[height=5.5cm,width=8.5cm]{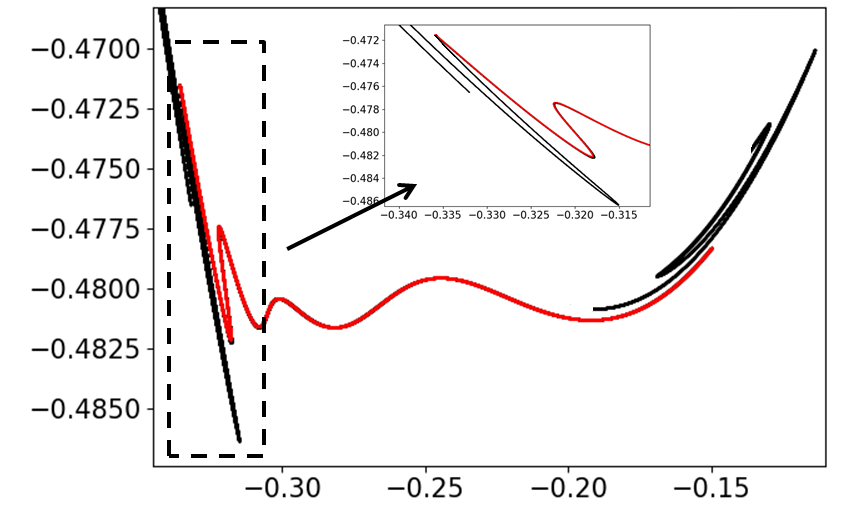}};
	\node[below of=russell, node distance=2.9cm, anchor=center,xshift=0.5cm] {{\Large{$x$}}};
	\node[left of=russell, node distance=4.6cm, anchor=center] {{{\Large $y$}}};
	\end{tikzpicture}
	\caption{Effect of the backward integration. Red curve is the one obtained satisfying the conditions I-IV, and the black curve is the union of its backward integration for 5 map iterations.}
	\label{backward_int}
\end{figure}

\section{Results: LCS computation for realistic RFP systems} \label{risultati}

As we  mentioned in the Introduction, one of the transport mechanisms at work in a fusion device is  chaotic transport. Such a mechanism has been often deemed the worst and it is present, at some extent, in all fusion configurations. For long time chaotic transport problems have been mostly  addressed by studying the associated Poincar\'e plot. But it gives only a partial description of the transport: scientists were used to think that, in the chaotic region, transport processes can be considered as random walks on the Poincar\'e surface (i.e. the surface used to draw the Poincar\'e plot). Then, arguments based on cantori\cite{AUBRY1983240,percival} (see section below) and ghost surfaces\cite{hudson_ghost} showed that this thought is a mistake. {Indeed}, random walk processes would not be able to sustain as high temperature gradients as those observed in chaotic regions even in experiments  \cite{evans2006,Sakakibara_2008,Suzuki2013}. In ref. \onlinecite{evans2006} the authors find that the applied magnetic perturbation increases the particle transport significantly more than the energy transport. These insights recall the work done in ref. \onlinecite{decastillo2011}, where the radial heat transport has been shown to be incompatible with the quasilinear diffusion model.  
{Such behaviour has been observed in all the magnetic configurations but, since  RFP transport is {believed to be} mainly due to chaos, we focus on such a configuration.  For many years it has been thought that the RFP, due to its intrinsic chaos, cannot sustain high temperature gradients. Such paradigm was shown to be false in the last decades {(RFX-mod\cite{sonato2003machine} observations\cite{Lorenzini,Gobbin_2013})}, but the physical mechanism behind the formation of such barriers still lacks {definitive} answers. Supported by the numerical achievement we show here, a further investigation of such techniques over experimental data would be very useful.}

In this section we are going to show how the LCSs  provide a  useful method to describe  transport, showing that it fits with another Lagrangian Descriptor (LD), i.e. the connection length, and that it is perfectly able to locate gradients of temperature.  The correspondence between LCSs and  temperature gradients is a strong promising result {to explain the experimental observations}.\\
In particular, exploiting the fact that energy transport is strongly anisotropic, we look for the LCSs of the magnetic field line system. As is well known \cite{digiannatale1,cary1983noncanonical,kruskal1952some,morozov1966structure,boozer1981plasma}, due to their solenoidal
nature, the field lines of a magnetic field in three-dimensional
space that does not vanish within the domain
of interest can be described at any fixed physical
time $t = \bar{t}$ as trajectories of a non-autonomous Hamiltonian system. The role of
time is played by a spatial coordinate taken to label the
points along a field line. The importance of this Hamiltonian formulation stems
from the fact that it establishes a direct connection between
magnetic configurations and dynamical systems {enabling the application of the techniques we use in this work to study  transport properties at a fixed time instant.}

\subsection{Realistic numerical cases}\label{model}
The magnetic fields here analysed have been obtained with MHD simulations performed with the Specyl code \cite{cappello_1996}. Specyl, successfully benchmarked with PIXIE3D code \cite{specyl_pixie,chacon2008},  solves the visco-resistivity MHD equations considering constant and uniform normalized density ($\rho = 1$ in the momentum equation). The
pressure term is  neglected: this approximation
is often used in dealing with macroscopic behaviour of systems characterized by strong current driven
activity (classic RFP feature). The equations written in dimensionless units are:
\begin{alignat}{2}\label{MHD_specyl_ini}
\partial_t {\bf v} + {\bf v} \cdot \nabla {\bf v} & =  {\bf j} \times {\bf B} + \nu  \nabla^2 {\bf v} \quad ,\\[0.15cm]
\partial_t {\bf B} & =  \nabla \times ( {\bf v } \times {\bf B} - \eta {\bf j} )  \quad , \\[0.15cm]
\nabla \times {\bf B} & =  {\bf j}  \quad ,\\[0.15cm]
\nabla \cdot {\bf B} & =  0 \quad . \label{MHD_specyl_end}
\end{alignat}

{Time-independent resistivity and viscosity profiles are prescribed accordingly to their experimental values}. In order to mimic a radial profile resulting from a Spitzer law, namely $\eta \propto T^{-3/2}$, the resistivity increases towards the edge according to: $\eta = \eta_0 (1 + 20 \, (r/a)^{10})$.  Numerical simulations are performed in cylindrical geometry with aspect ratio $R_0/a = 4$, and we take $a = 1$ as the cylinder radius. The
edge radial magnetic field is either zero (the so-called 'ideal
conducting wall'), or helically modulated through a magnetic
perturbation (MP) with {chosen} poloidal and axial wave number ($m_{MP}, n_{MP}$).  The numerical setup is as follows. The dimensionless central transport parameters resistivity ($\eta = S^{-1}$) and viscosity ($\nu = M^{-1}$) are $10^{-6}$ and $10^{-4}$ respectively: they were chosen to enter a quasi-cyclical  {regime} which is very similar to what is observed in  RFX-mod experiments. 
As for the boundary conditions, MPs are applied in order to attain experimental-like cycles of Quasi Single Helicity (QSH) states with a dominant MHD mode which has the same twist as the applied MP\cite{bonfiglio2013,Veranda_2019}. 
 During the cycle of QSH formation, the system features wide regions with magnetic chaos.
In this section we are going to analyse magnetic fields coming from two kinds of simulations. 
The simulations differ only with respect to the applied MP boundary condition (and thus the dominant MHD mode), either the non-resonant helical twist $m=1_{MP}, n_{MP} = -6$ or $m=1_{MP}, n_{MP} = -7$ that is resonant {(or marginally resonant)}.  In both cases  amplitude $\text{MP}_\% = 2\%$ {which is close to the experimental mean edge amplitude}.\\
The field line trajectories have been obtained with the field line tracing code NEMATO  (presented in Ref. \onlinecite{nemato} and numerically benchmarked in Ref. \onlinecite{ciaccio_veranda}). {In all the LCS computations, we used the auxiliary grid technique (sec \ref{auxiliary_grid_section}), the backward integration (sec \ref{back_int}) and $1000$ i.c. in both radial ($r \in [0, 0.7]$) and poloidal direction ($\theta \in [0, 2\pi]$). Such a choice is a good compromise between computational time and result accuracy. The reader could see that the i.c. discretization is much denser in radial than in poloidal direction; this choice is related to the knowledge that, for the system we analyse, the structures are much more localized in radial rather than in poloidal direction}. {It is worth to mention that most of the computational time is due to the computation of the i.c. trajectories}.

\subsection{LCS and connection length} \label{connection_length_section}
The connection length (CL) can be thought as a Lagrangian descriptor of the field line motion since its value is associated to the  trajectory of each field line. Commonly it is the average, between forward (usually the toroidal angle increases) and backward (usually  the toroidal angle decreases)  computation of the arc-length that a field line has to tread till reaching the wall. In this section, for analogy with the repelling LCSs {that take into account the forward dynamics}, we just take the arc-length for the forward dynamics. Moreover, due to the  peculiarities of the RFP magnetic topology featuring an edge transport barrier given by a chain of $m=0$ islands at the reversal surface\cite{spizzo_PRL}, we record the arc-length needed to reach the radial position $r/a = 0.8$.\\
Figure \ref{LCS+lc} shows {(on a poloidal section of the simulation domain) the map of computed LCSs and CL. The white spot on the plot correspond to region occupied by regular KAM surfaces, while black spots are region where field lines quickly escape. The LCS (white curves) appear to separate regions with different colours of CL map: LCS are in agreement with the CL}. The magnetic field refers to the onset of a QSH RFP state with dominant component of the magnetic field Fourier spectrum $m=1,n=-6$ with $m, n$ poloidal and axial mode number. In order to give an idea of the physical time-scales, let us consider an electron with an energy of 700 eV, i.e. the typical RFX-QSH state value.  With this energy an electron travels a length $L = 10^5 m$  in a time interval $\Delta t \simeq 10^4 \tau_A$ (with $\tau_A$ Alfvén time), that is much higher than the time scales on which the magnetic field changes. {Moreover the mean free path is estimated to be around 10 toroidal loops\cite{Veranda_2017}.}\\
The picture highlights the importance of such LCSs in ruling the system dynamics.  LCSs have to be thought as {\itshape dynamical} transport barriers: this means that, given a certain LCS, it  would be wrong to suppose that particles-field lines living inside a certain region delimited by an LCS cannot move to another part of the domain.  {Any region moves according to the motion of LCS (moving under the flow map) that encloses the region itself.}
To check if computed LCSs really act as transport barriers,  previous works\cite{digiannatale2,rubino2015detection,borgogno2011barriers,Veranda_2017}  evolved the points separated by LCSs constructing a kind of "reduced" Poincaré plot. This approach only works for regions "separated" by strong cantori (as we explain in the next section). In general, in order to check if an LCS acts as a dynamical (that is Lagrangian) transport barrier, one should compute at each "time" the  position of  LCS and the position of points evolving under the action of the flow map. \\
However, it is still possible that LCSs separate regions having a reduced  transport even from the Eulerian point of view: this happen when LCSs give rise to an accumulation of ribbons. The LCSs accumulation is a signature of the Eulerian reduced transport between regions that results  in a drop of the connection length as can be seen in several regions of fig \ref{LCS+lc} divided by such bundles of LCSs. Such a behaviour will be better investigated and explained in the following subsection. \\
Concluding, all the LCSs in fig \ref{LCS+lc} play a crucial role for the field line motion. Some of them can almost act as {\itshape static-Eulerian} barriers and the other ones are still important because they describe how and how fast the transport process develops.

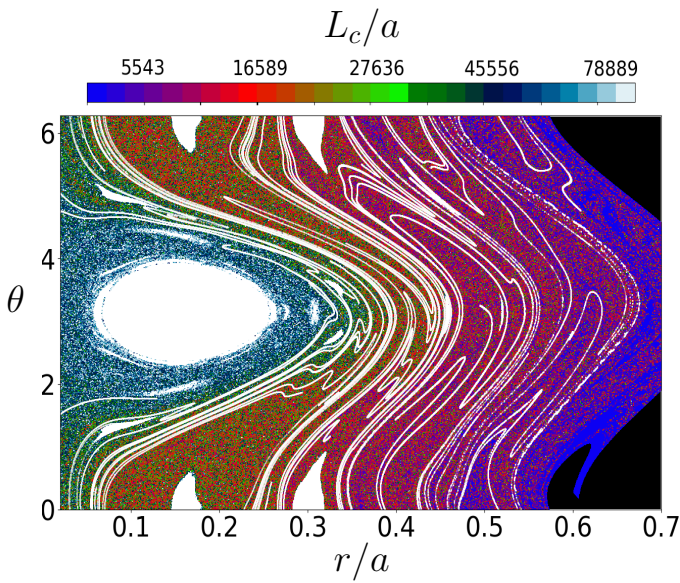
\begin{figure}
	\begin{tikzpicture}[trim left=-4.5cm]
	\node[node distance=1cm] (russell) at (0,0)
	{\includegraphics[height=6.5cm,width=8.5cm]{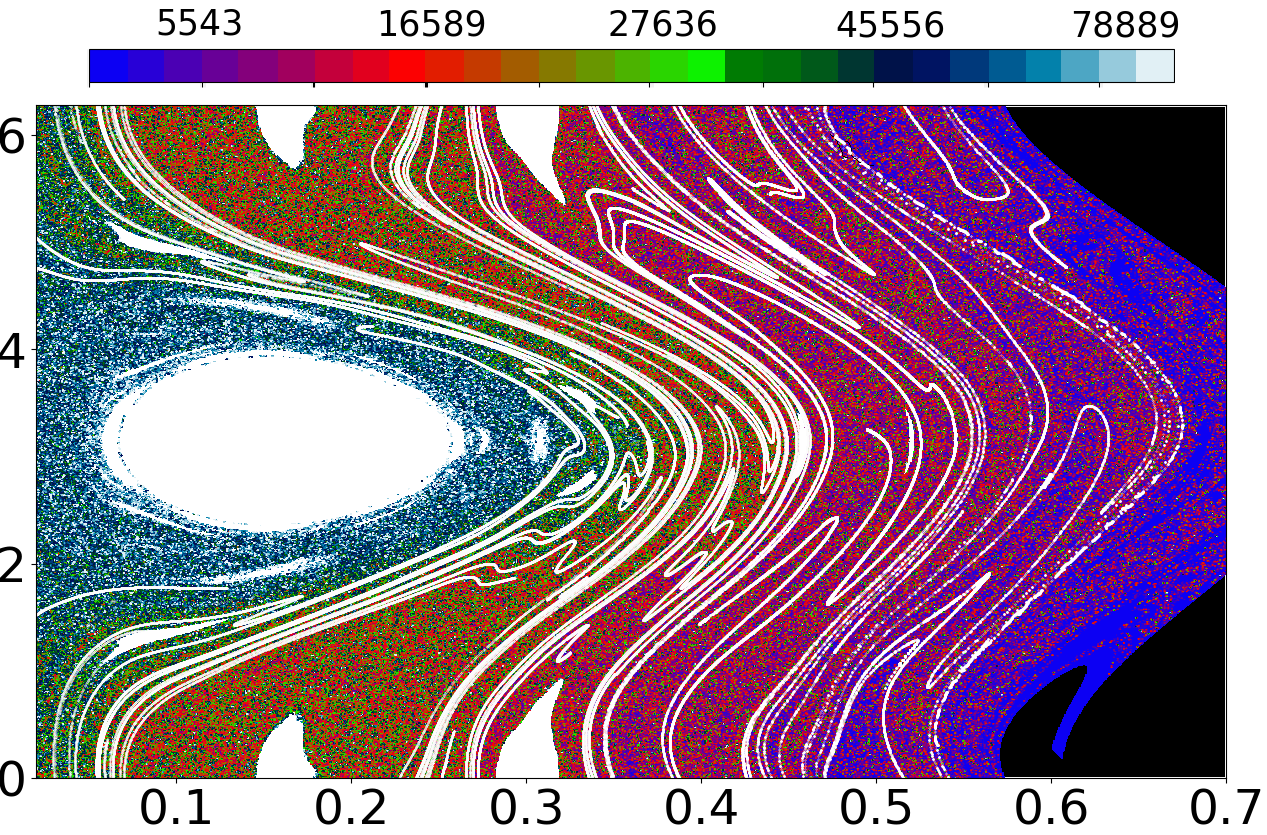}};
	\node[below of=russell, node distance=3.5cm, anchor=center] {\Large{$r/a$}};
	\node[left of=russell, node distance=4.6cm, anchor=center] {{\Large $\theta$}};
	\node[above of=russell, node distance=3.6cm, anchor=center] {{\Large $L_c/a$}};
	\end{tikzpicture}
	\caption{Repelling LCSs (white curves) and connection length during a QSH state with dominant magnetic mode $m=1,\; n=6$ (poloidal and toroidal mode number respectively). Here the connection length is a dimensionless parameter since it is normalized to the cylinder radius.}
	\label{LCS+lc}
\end{figure}

\subsection{LCS and cantori}

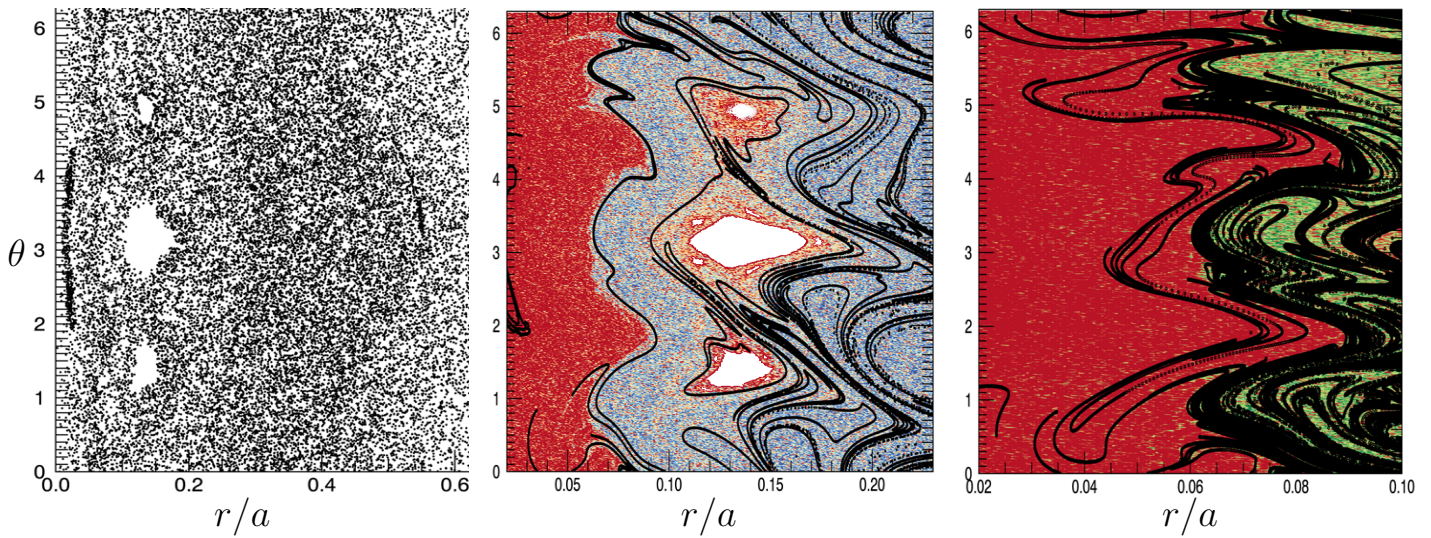
\begin{figure*}[htp]
	\begin{tikzpicture}[trim left=-2.5cm]
	\node[node distance=1cm] (russell) at (0,0)
	{\includegraphics[height=6.5cm,width=6cm]{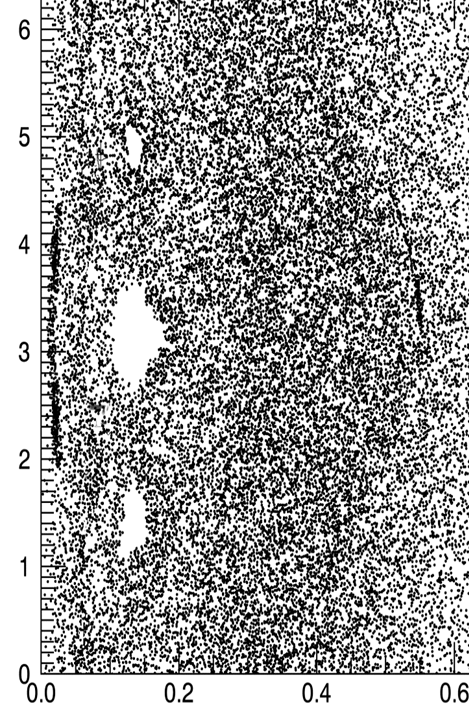}};
	\node[below of=russell, node distance=3.5cm, anchor=center] {\Large{$r/a$}};
	\node[left of=russell, node distance=3cm, anchor=center] {{\Large $\theta$}};
	\node[right of= russell,node distance=6.2cm, anchor=center ] (luigio) 
	{\includegraphics[height=6.5cm,width=6cm]{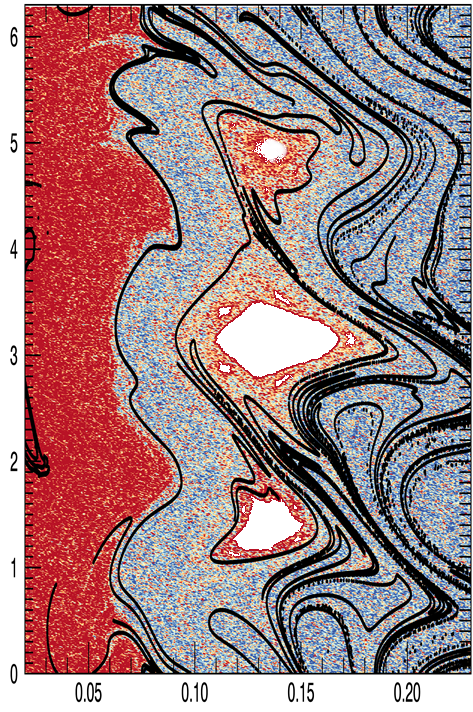}};
	\node[below of=luigio, node distance=3.5cm, anchor=center] {\Large{$r/a$}};
	\node[right of= luigio,node distance=6.4cm, anchor=center ] (papo) 
	{\includegraphics[height=6.5cm,width=6cm]{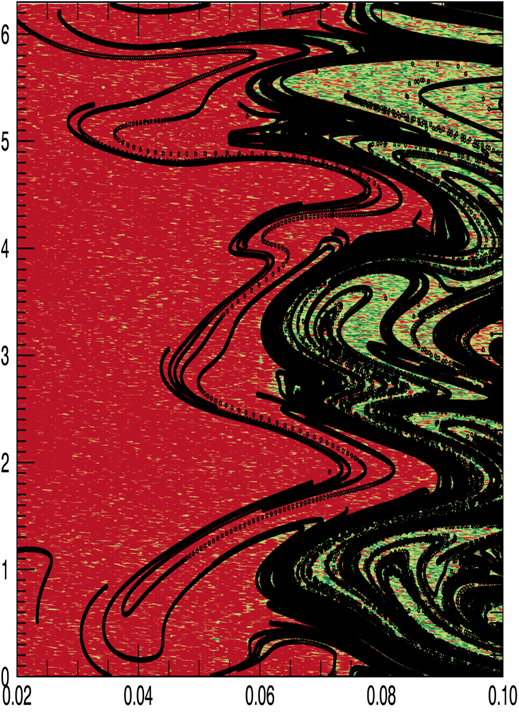}};
	\node[below of=papo, node distance=3.5cm, anchor=center] {\Large{$r/a$}};
	\end{tikzpicture}
	\caption{Field line transport during a MH state of the simulation  $m_{MP}=1, n_{MP}=-7$. Left panel: Poincaré plot built with 20 x 20 orbits initially distributed in $r \in [0.001, 0.6]$ and $\theta \in [0, 2 \pi]$. Central panel: connection length  and LCSs  focusing on the region $r \in [0.02,0.23]$. Right panel: connection length  and LCSs (backward integrated for 10 axial loops)  focusing on the region $r \in [0.02,0.1]$}
	\label{LCS_lc_MH}
\end{figure*}
The possibility for LCSs to act as static transport barriers is linked to their relation with Cantor sets \cite{Lichtenberg,mackay1984stochasticity}. 
KAM theory\cite{KAM_esist} guarantees the existence of invariant tori with sufficiently irrational frequency vectors in Hamiltonian systems sufficiently close to integrability. Such invariant tori disappear when the system is far from being integrable. The existence of remnants of
these tori after they have been destroyed was first deduced by
Aubry and Percival using variational arguments \cite{AUBRY1983240,percival}: such mathematical objects acting as transport barriers  are called cantori.\\
The importance of cantori was confirmed in \cite{MACKAY198455} when it was discovered that the flux of phase space area is locally minimal on cantori. The transport through cantori depends on "how much" the system is locally far from being integrable\cite{Escande_1985}, according to the rule $\Gamma \sim (s - s_c)^\nu$ where $\Gamma$ is the flux, $s, s_c$ are respectively the parameter controlling the chaos amplitude and the threshold for KAM torus considered, and finally $\nu$ is a critical exponent. Thus, an Hamiltonian system can have static (or better Eulerian) transport barriers only if cantori are in the system.
We have seen that there are circumstances in which the LCSs can help to locate such cantori and then such LCSs act as barriers. \\
In the left panel of fig. \ref{LCS_lc_MH} {we show} a Poincar\'e plot for a {regime} characterized by the interaction of a wide spectrum of MHD instabilities naturally resulting in a stochastic behaviour of field lines, the so-called MH regime. \\
The chaotic domain appears fully connected and, apparently, a  field line can move in the whole chaotic region as if no transport barrier were present. Nevertheless, drawing the LCSs and the connection length, focusing on the radial region $r/a \in [0. , \; 0.25]$, one has a full understanding about the transport process. This is shown in the central panel of fig. \ref{LCS_lc_MH}.\\
The first thing to be noted is the fact that near the islands (that clearly have an infinite CL) there are points with quite high value of the CL. Such a behaviour is remarked by the LCSs shape: near the islands the LCSs exhibit the classical lobes that one expect from the stable manifolds, in particular the LCSs are able to track the manifolds till the first homoclinic intersections. For this reason the points included in these lobes have to go further than the other ones before reaching the edge. On the opposite, on the right of the islands the LCSs have a much more intricate shape enclosing broader regions. Such a behaviour is a strong evidence of the fact that in this region mixing phenomena happen very fast, leading to higher transport and lower CL. \\
The other  thing to be noted is  that there is a sudden drop of the CL in the region around $r/a = 0.06$ where the colour plot suddenly changes  from red-yellow to blue-yellow. This means that there is an object reducing the transport. We can be sure that in that region a KAM surface cannot exist because the internal region has a finite CL. Moreover, yellow points can be detected in the region $r/a < 0.012$, indicating that points exists from where magnetic field lines can quickly escape, as if a preferential channel were present\cite{Spizzo_2007}. This is a clear evidence of a Cantor set that locates where the CL drops. Such behaviour is  remarked even by the LCS shape. Around the Cantor set, the LCS disposes almost along it with only a very small penetration inside the region delimited by the cantorus. \\
To check how a LCS "feels" the presence of a Cantor set, we back-integrated them for 10 toroidal loops. The result is in the right panel of fig. \ref{LCS_lc_MH} that highlights the radial region $r/a \in [0.02,0.1]$. The backward integration of the LCS, as explained in section \ref{back_int}, gives rise to a stretching of the curve itself. But the curve does not stretch uniformly in space: most of the spreading happens inside the chaotic region where the LCS lives and, even after the stretching, only a little part of the LCS is able to penetrate inside the red region. This penetration has to happen, otherwise the two regions would be completely separated by a KAM surface; however, such a penetration is very low and it happens thanks to the gaps of the Cantor set. Thus, it is possible to see how the LCS tends to accumulate around the Cantor set after its back-integration.

\subsection{Comparison with temperature distributions}

As already mentioned, heat transport in magnetized plasmas is inherently anisotropic 
with $\chi_\parallel/\chi_\perp$ exceeding $10^{10}$ in fusion {configurations} \cite{braginski_anisotropia}.
For this reason the topology of magnetic field lines strongly affects the resulting temperature
profiles. Moreover, the impact of the magnetic field  on the  temperature 
profile is also due to the fact that different fields give rise to different Ohmic
heat sources,  $S = \eta j^2$.\\
To compute the temperature {distribution associated with cases at hand} we used the MHD anisotropic heat transport equation solver T3D \cite{t3d}. The code solves the equation
\begin{equation}
	\frac{\partial T}{\partial t} - \chi_\parallel \nabla^2_\parallel T -\chi_\perp \nabla^2_\perp T = S = \eta j^2
\label{heat_transp}
\end{equation} 
using a semi-Lagrangian approach which transforms it in an integro-differential equation that can be solved through the Greens's function formalism. The integration of the Green's function is performed along magnetic field lines avoiding the perpendicular transport pollution by high parallel transport which is a limitation of numerical solutions of Eq.\ref{heat_transp} on a grid. {The equation is solved taking into account normalized coefficients, so the only important parameter is the anisotropy ratio, $\chi_\parallel/\chi_\perp$}.
In the simulations that we show,  $\chi_\parallel/\chi_\perp = 10^7$ is constant (i.e. not self-consistent) and uniform. {The heat source is given by $\eta j^2$ :  due to the fact that RFP plasmas do not require additional heating systems, the Ohmic term is a good approximation of the heating power.} It is important to remark that equation \ref{heat_transp} is not coupled with the MHD visco-resistive model of Sec \ref{model}. The code T3D solves  equation \ref{heat_transp} with a given (not evolving) magnetic field obtained by solving the visco-resistive model (eqns. \ref{MHD_specyl_ini}-\ref{MHD_specyl_end}).

We aim at showing that even chaotic magnetic fields can sustain relevant temperature gradients and that such gradients are perfectly located by LCSs. {Such correspondence can be found for states in which chaos seems to be more (MH) or less (QSH) widespread.}
We expect LCSs to offer a more precise description of temperature gradients than CL. This is basically due to two reasons. The first one is related to the fact that the connection length approach fails when there are chaotic regions separated by KAM surfaces: in this situation the connection length would be infinite because, due to the presence of a KAM surface, field lines never reach the  edge. A similar argument holds when strong cantori live in the chaos: the cantori would allow only a small flux through them and this tends to uniform the CL that would result high for all the points. In general, the CL approach could fail when there are several chaotic regions  poorly connected, that is with a low flux among them.\\
{The second reason for LCSs being more effective than the CL approach to describe temperature barriers  is related to the fact that, when there is a local heat deposition, the LCS distributes most of that heat along themselves: the heat transfer along the LCS itself is much larger than the heat transfer across the LCS because of the high anisotropy.}\\
The reader should keep in mind that in this kind of non-autonomous periodic systems, the role of repelling and attracting structures is exactly the same because, as previously said, the most attracting structures can be found as repelling for the backward dynamics. 
This is respected by the T3D code which computes the parallel transport for both forward and backward dynamics: the integration on the arch-length, $s$, is performed between $s \in [\text{Tr}^-,\text{Tr}^+]$ where $\text{Tr}^-,\text{Tr}^+$ represents a threshold on  the arc-length because, for practical reasons, the integration cannot be done along the whole field line length (i.e. infinity) \cite{t3d}. 
{{We now} analyse {two} magnetic cases corresponding to  different amplitudes of the dominant mode ($m=1, n=7$): QSH and  QSH crash (MH).

{The Poincar\'e map of the first case,  shown in fig \ref{poinc_temp_QSH_n7}, shows a core region with conserved KAM surfaces embedded in a rather homogeneous chaotic domain, which could hardly be imagined to host a transport barrier}. 
\begin{figure}[htp]
	\begin{tikzpicture}[trim left=-4.cm]
	\node[node distance=1cm,yshift=3cm] (russell) 
	{\includegraphics[height=5.5cm,width=7cm]{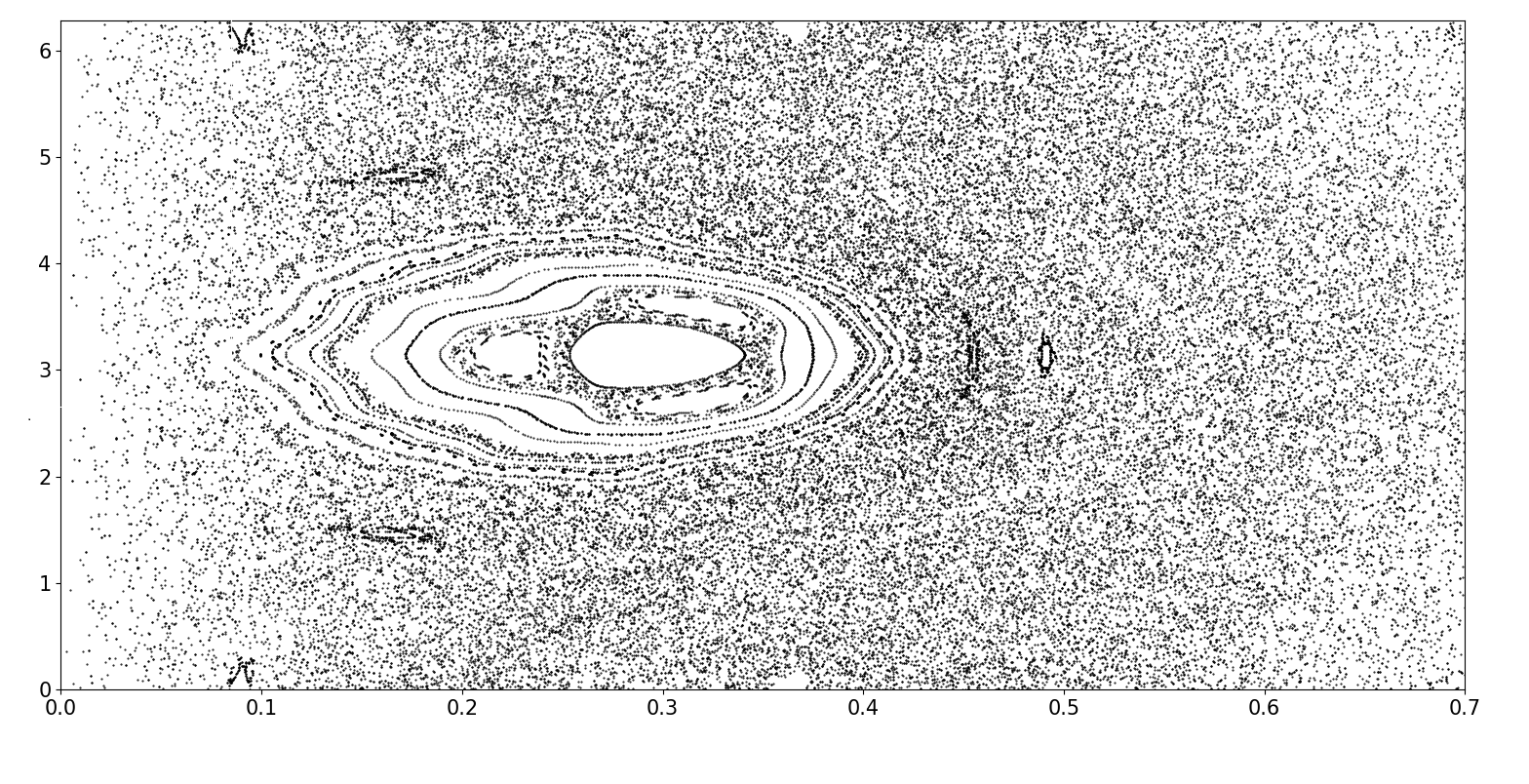}};
	\node[below of=russell, node distance=2.8cm, anchor=center] {\Large{$r/a$}};
	\node[left of=russell, node distance=3.8cm, anchor=center] {{\Large $\theta$}};
	\end{tikzpicture}
	\caption{Poincaré plot of the QSH state with dominant helicity $m=1, n=-7$, with $m, n$ poloidal and toroidal mode number respectively.}
	\label{poinc_temp_QSH_n7}
\end{figure}
However, the temperature plot, showed together the LCSs in figure \ref{LCS_temp_QSH_n7}, exhibits  high  temperature gradients inside  the chaotic region underling how even inside a fully chaotic region temperature gradients can be sustained.\\}
\begin{figure*}[htp]
	\begin{tikzpicture}[trim left=-4.7cm]
	\node[node distance=1cm] (russell) at (0,0)
	{\includegraphics[height=6.5cm,width=8.cm]{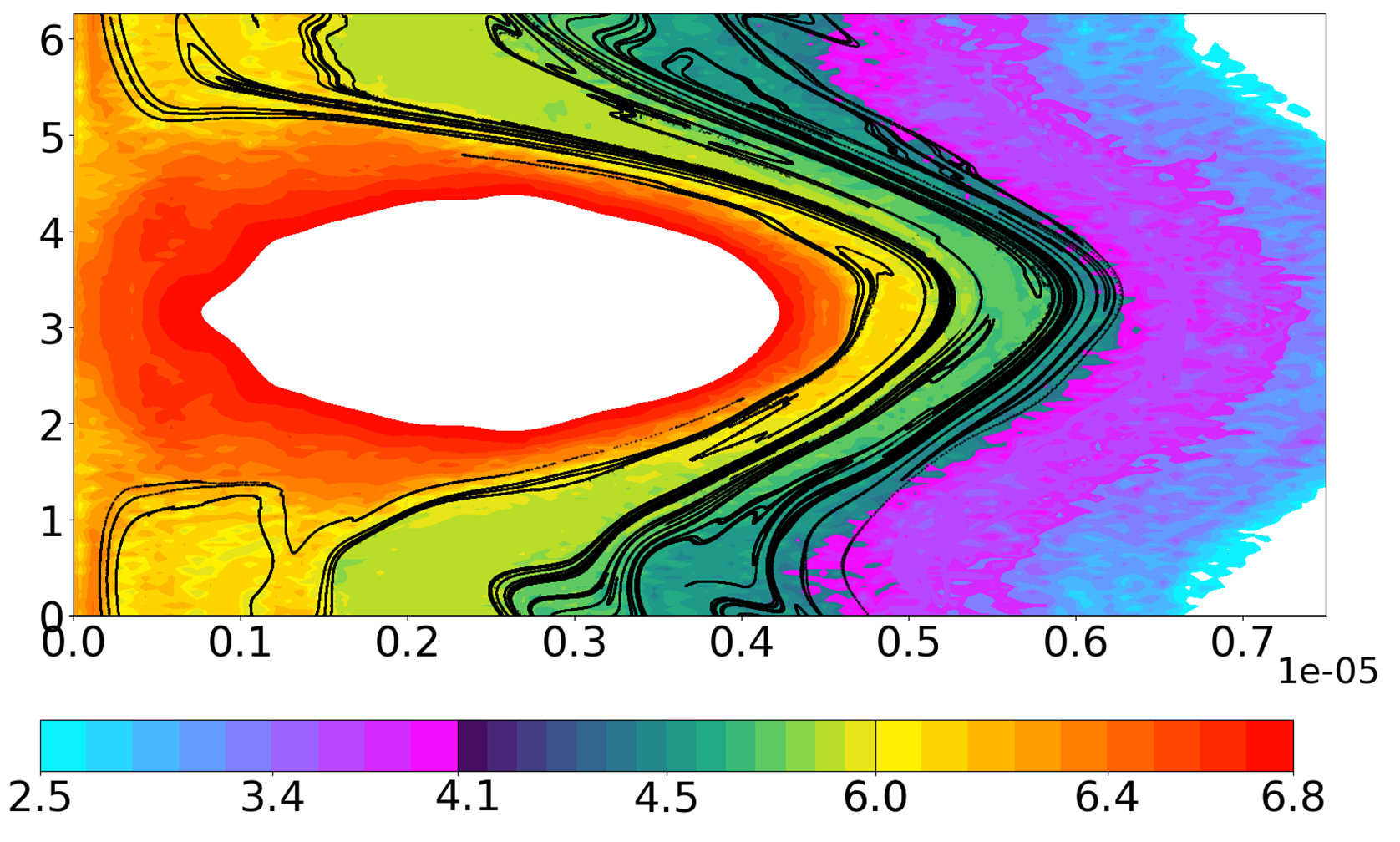}};
	\node[below of=russell, node distance=3.4cm, anchor=center,xshift=-2mm] {\Large{$r/a$}};
	\node[above of=russell, node distance=3.4cm, anchor=center,xshift=3cm] {{a.u.}};
	\node[left of=russell, node distance=4.3cm, anchor=center,yshift=8mm] {{\Large $\theta$}};
	\node[right of= russell,node distance=9.5cm, anchor=center,yshift=1mm ] (luigio) 
	{\includegraphics[height=6.5cm,width=8.cm]{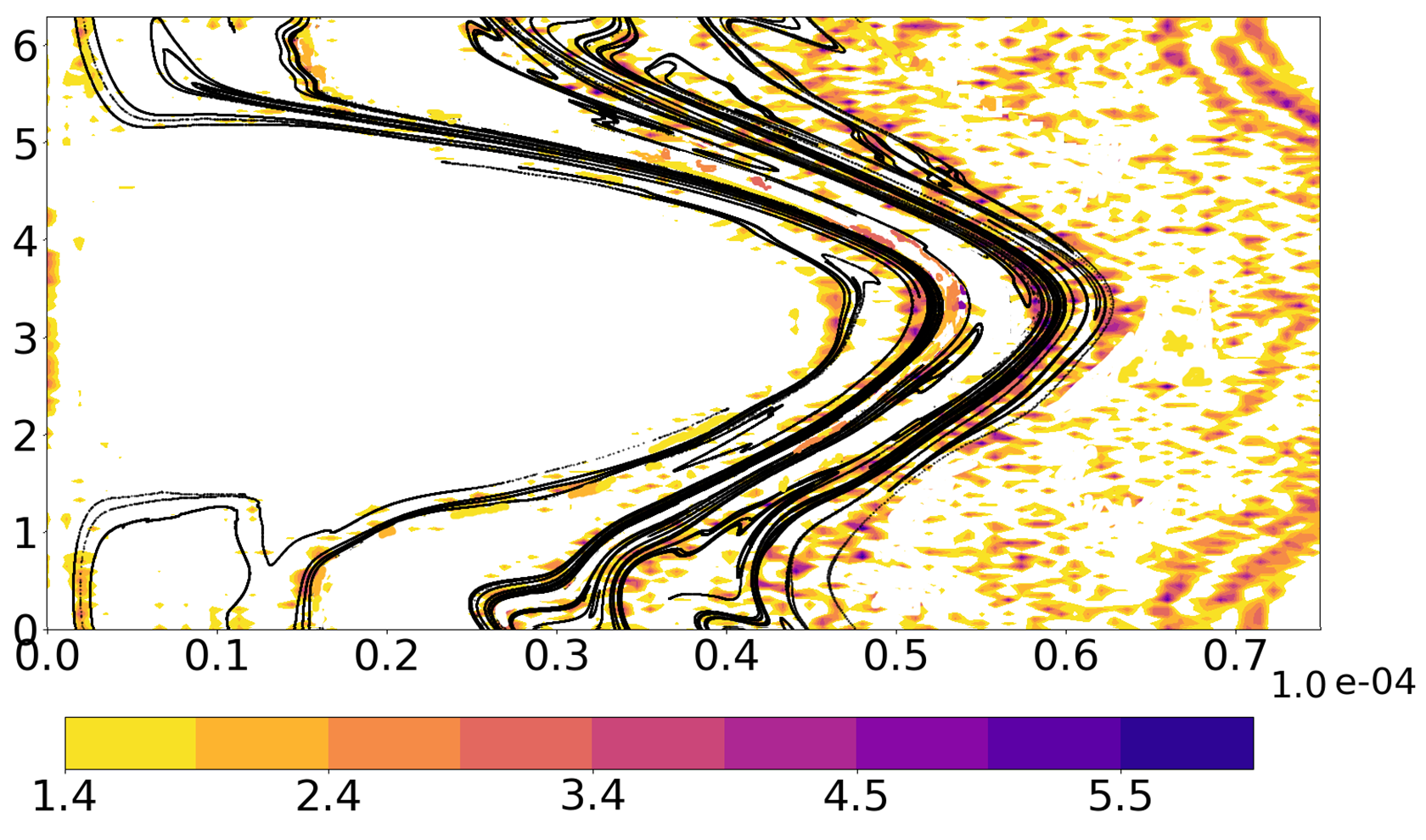}};
	\node[left of=luigio, node distance=4.5cm, anchor=center,yshift=6mm] {{\Large $\theta$}};
	\node[below of=luigio, node distance=3.5cm, anchor=center] {\Large{$r/a$}};
	\node[above of=luigio, node distance=3.3cm, anchor=center,xshift=3cm] {{a.u.}};
	\end{tikzpicture}
	\caption{LCSs for the QSH state with dominant helicity $m=1, n= -7$,with $m, n$ poloidal and toroidal mode number respectively. Left panel: comparison with temperature, the white region is the hottest one ($T>6.8 10^{-5}$) and it corresponds to the plasma helical core. Right panel: comparison with the temperature gradient, in the white region the gradient is smaller that $1.4 10^{-4} $. {The numerical values are given in arbitrary units.}}
	\label{LCS_temp_QSH_n7}
\end{figure*}
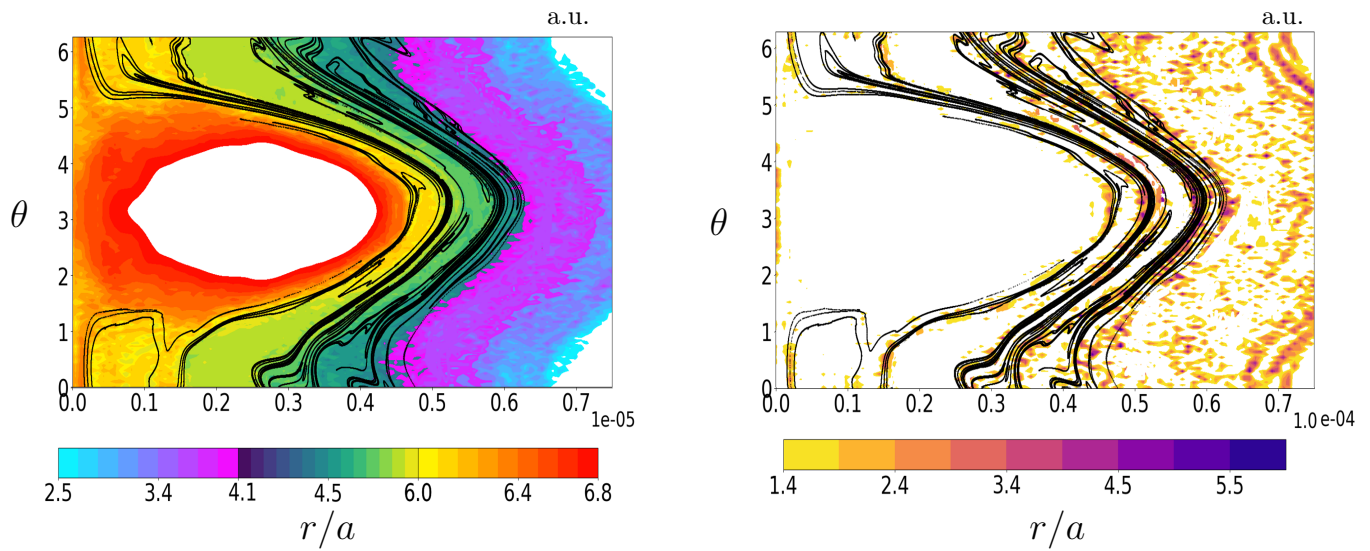
The left panel shows both temperature and LCSs. It is immediate to catch the correlation between both: LCSs  separate regions featuring different temperature values. {As explained in section \ref{section3}, a lot of LCS may be extracted} but, taking only those with high repulsion rate (eq. \ref{mean_repulsion}) according to the ordering procedure described in section \ref{section3}, we  select the main LCSs.\\
In the right  panel the comparison between LCSs and temperature gradient is explicit. We can see how LCSs lie on the highest value temperature gradient (in module).  Concluding, figure \ref{LCS_temp_QSH_n7} shows how deep is the correlation between LCSs and temperature, indicating that such ruling patterns are the key ingredient for the formation of temperature gradients. {If such ruling patterns did not exist and the motion of field lines on the $r-\theta$ plane obeyed to a uniform random walk, then the temperature, and so the temperature gradient, would not be  high since the anisotropy effects would strongly decrease. The information of the parallel direction would be mixed up with the perpendicular one resulting in a practical increase of $\chi_\perp$ and then to a decrease of temperature.} 

The magnetic field analysed in Fig.\ref{LCS_temp_QSH_n7} corresponds to a QSH state where a single Fourier harmonic  is dominant in the magnetic spectrum. In this case, well defined temperature gradients, and so LCSs, can be found. However, the correspondence between LCSs and temperature can be found even for fields featuring larger chaotic region. \\
The case we analyse in the following refers to the MH state already described in section {\bf V-C}.
As we said, looking at the Poincar\'e in figure \ref{LCS_lc_MH}, the magnetic chaos extends to almost the whole domain and, according to old thinking, temperature gradients are not supposed to exist.\\
Nevertheless, thanks to the fact that LCSs  rule the field line transport process, it is still possible to observe high gradients (comparable to the QSH case). Being a full MH magnetic field, this time the LCSs are much more intricate and convoluted with respect to the cases we showed above, respectively onset-QSH (figure \ref{LCS+lc}) and QSH (figure \ref{LCS_temp_QSH_n7})  where there is a single Fourier component of the magnetic field much bigger than the others. Such intricate behaviour of LCSs reflects on the shape of the temperature distribution that results much more {oscillating, in both radial and poloidal directions,} with respect to QSH and onset QSH states. For this case, due to the strong LCSs asymmetry with respect to  $\theta=\pi$, the Lagrangian-regions enclosed and transported by LCSs are much more localized in $\theta$. Thus, in order to fully describe the skeleton of the dynamics we need to draw both repelling and attractive LCSs. As shown in Ref \onlinecite{digiannatale1}, due to symmetry property of the system, the attractive LCSs can be drawn as mirror images of the repulsive LCSs with respect to $\theta = \pi$. Temperature map and LCSs for this MH case are shown in fig. \ref{MH_LCS_temp}.\\
As mentioned, the LCSs look much more intricate than in the previous cases. {Even though their convoluted shape indicates  strong temperature oscillation, they still } prevent a diffusive process. The areas enclosed by the LCSs are the quantities  giving indication about how strong the transport is: the LCSs move and map into each other {under the action of the flow map $\Phi$} allowing areas to move around the domain. As much as the LCSs are convoluted, as much the regions enclosed by LCS tend to move rapidly around the domain increasing  transport. {In conclusion, even for this very chaotic magnetic field it is possible to find temperature gradients and LCSs behave accordingly to the gradient shapes. This is possible thanks to the fact that LCSs govern the transport processes according to Fig \ref{MH_LCS_temp}}. 

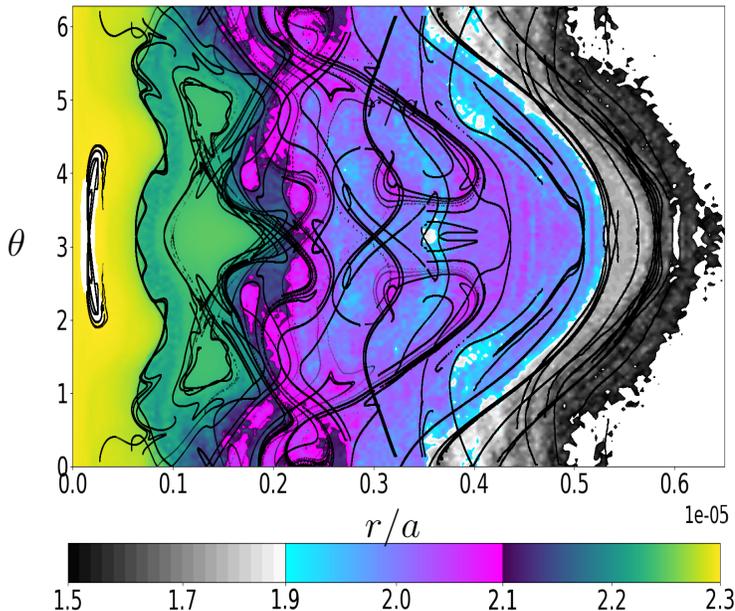
\begin{figure}[htp]
	\begin{tikzpicture}[trim left=-4.5cm]
	\node[node distance=1cm,yshift=3cm] (russell) 
	{\includegraphics[height=8.5cm,width=9.5cm]{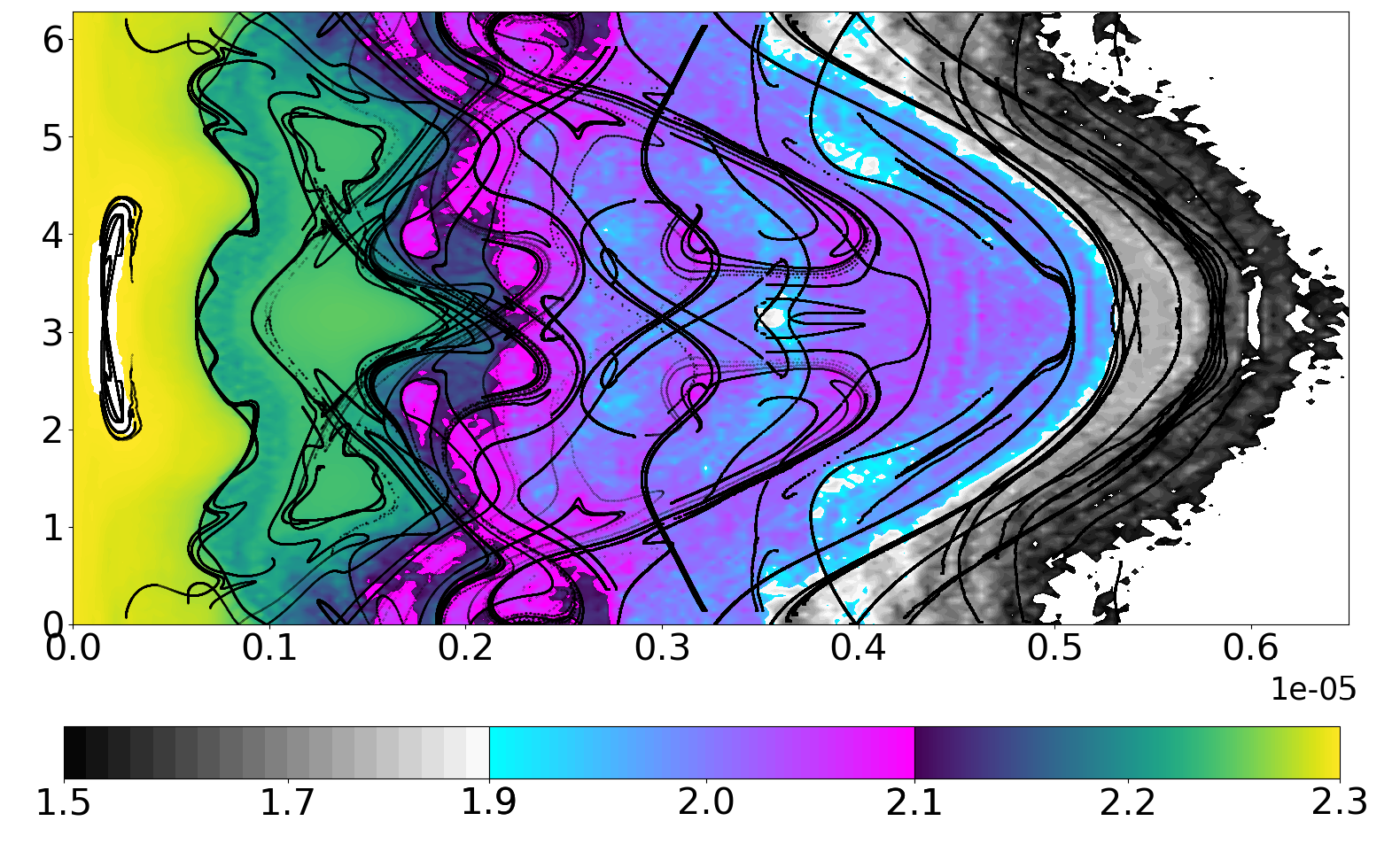}};
	\node[below of=russell, node distance=2.8cm, anchor=center] {\Large{$r/a$}};
	\node[above of=russell, node distance=2.8cm, anchor=center] {\Large{$r/a$}};
	\node[left of=russell, node distance=5.cm, anchor=center,yshift=1cm] {{\Large $\theta$}};
	\end{tikzpicture}
	\caption{Comparison between LCSs and temperature map. The magnetic field refers to a MH state that follows a QSH crash with dominant helicity $m=1, n= -7$, with $m, n$ poloidal and toroidal mode number respectively. The numerical values are given in arbitrary units.  }
	\label{MH_LCS_temp}
\end{figure}
{We conclude this section with a consideration concerning how LCSs, or better the mathematical objects they try to track, help to sustain a good temperature profile and what is the limit of our physical model with respect to the experimental evidence.} \\
{In the numerical simulations we describe, the temperature is still far from the experimental shape, which exhibit a sudden drop of the temperature in a localized region.  As shown in Ref. \onlinecite{dhaesler}, it is possible to solve the temperature equation in  flux coordinate geometry taking into account only the dominant magnetic Fourier component. This is the so called 1.5D approach. In this frame, where {\itshape ipso facto} chaotic transport does not exist, solving the transport equation corresponds to considering the flux surfaces as iso-thermal. 
	The  just described situation can be considered as a limit for a chaotic system: LCSs (or better the objects they track) help  preserving an "order" in such a way to avoid that strong radial jump of magnetic field lines can lead to a flattening of temperature due to the high value of $\chi_\parallel$; thus LCSs definitively keep a separation between perpendicular (low) and parallel (high) diffusion. If LCSs were not in the system, different regions would mix so much that it would not be  possible  developing temperature gradients as those in the 1.5D framework.\\
	In order to recover the experimental shape of the temperature, as shown in Ref.  \onlinecite{Martines_2011, Gobbin_2013} that assume well defined flux surfaces, it is necessary to have a drop of $\chi_\perp$ at a certain flux surface (an LCS for a full 3D transport model): a non-uniform, hollow, thermal coefficient is required.  To take into account this phenomenon, one should compute local parameters starting from macroscale transport but it is beyond the scope of this paper. As an example concerning the neglected phenomena, we can simply recall (being aware that is not enough to produce the drop one needs) that, in the Braginskij's formulation, the
	$\chi_\perp$ is linked to the temperature through a law as $\chi_\perp \propto T_e^{-1/2}$ thus a local increase of the temperature on the LCS leads to a decrease of   $\chi_\perp$. 
	This happens quite often because along the LCS  local peaks of heat
	deposition can arise and such heat 
	deposition is then distributed along LCS
	itself.}
\section{Summary, final remarks and future prospectives}

The focus of this paper is Lagrangian Coherent Structures that are the skeletons  governing the motion of the system, i.e. the magnetic field lines for our studies. The paper first presents a {Python} numerical tool we developed to compute  such structures and then it focuses on the physical application to numerical-realistic fusion plasma configurations. \\
The tool description focuses on three peculiar aspects. We have shown that it can work for a general geometry, a crucial aspect in fusion plasmas where most  quantities are described in flux surface geometry, i.e. a curvilinear geometry. Then we explained how the auxiliary grid and the backward integration improve the calculation. The first one allows a better computation of the deformation gradient and the related quantities (Fig \ref{auxiliary_grid_vp}), the latter allows for the extraction of longer and more intricate curves (Fig \ref{backward_int}) that describe better the dynamics.

The second part of the paper concentrates on physical applications of the LCS to MHD numerical simulations of the reversed-field pinch configuration.  We have shown that there is a remarkable agreement between the LCSs and another Lagrangian Descriptor, i.e. the connection length.  We showed that LCSs {can} be thought as dynamical transport barriers: {they move under the dynamics and the regions enclosed by the LCSs do the same, coherently with their boundary, i.e. the LCSs themselves.}
We showed that sometimes LCSs can be considered even as barriers in the classical term, i.e. under an Eulerian point of view. This happen when the system features cantori: LCSs tend to align along such cantori that have a low transport through them. Such cantori can develop even in a fully chaotic situation (MH state)  {locally lowering} the transport and allowing for gradients of physical quantities.

Finally, we focused on temperature transport equation and on the comparison of its solutions with the patterns described by LCSs. We have shown that even a chaotic magnetic field can feature significant temperature  gradients and then we  showed that the LCSs do account for this behaviour. The gradients can be sustained inside a  chaotic region thanks to the presence of LCSs who represent well defined structures ruling the motion of the field lines. 
The analyses showed that LCSs perfectly locate regions with largest temperature gradients. We  showed that the dynamics inside
chaotic regions cannot be simply represented by a random
walk because strong patterns rule the motion: if the motion of field lines on the $r-\theta$ plane obeyed
to a random walk then the temperature gradient would
 not be high since the anisotropy effects would strongly
decrease. As a future work, we plan to do the same analysis using magnetic field coming from reconstruction\cite{zanca} of experimental measurements performed in the  RFX-mod \cite{sonato2003machine,marrelli2007magnetic} device and using  tokamak and stellarator fields (both numerical and experimental).

All the analyses presented here take into account magnetic field line trajectories and thus they involve non-autonomous 2.5D systems. However, the LCS method is generally applicable 
to the full particle dynamics in 3D coordinate space
and by employing e.g., the exact particle Hamiltonian in
time varying electromagnetic fields. As further investigation beyond the field line analysis described here, we are considering the possibility to include effects related to micro-turbulence to better describe the tokamak physics. We plan on combining the chaotic-like transport with the  motion of test particle gyrocenters undergoing, among the other drifts, the effect of ${\bf E} \times {\bf B}$ advection in electrostatic turbulence, similarly to \onlinecite{Padberg_2007}.

\section{ACKNOWLEDGMENTS}

The authors thank Fabio Sattin for discussions {concerning the anisotropy effects for random processes and Dominique Escande for stimulating fruitful discussions concerning the Cantor sets properties}. The authors want to thank {Xavier Garbet and Andrea Garofalo for showing interest in the topic of this work.} The authors acknowledge use of the computational resources provided by the EUROfusion High Performance Computer (Marconi-Fusion) through the EUROfusion project named 'PIXIE3D', which includes the  studies described in section \ref{risultati}.


\appendix
\section{Numerical computation of the third condition} \label{appendiceA}

Here, in order to simplify the notation, $\boldsymbol{\xi}, \;\lambda$ stand for $ \boldsymbol{\xi}_{max}, \; \lambda_{max}$.  Mathematically speaking, being $\boldsymbol{\xi} $ computed as in \ref{eigenvectors}  $\boldsymbol{\xi} = \xi^i\, {\bf e}_i$, the computation of \ref{numericallyaproblem} is simply $ \boldsymbol{\xi} \cdot \nabla \lambda = \xi^i\; \lambda_j {\bf e}_i \cdot {\bf e}^j = \xi^i\; \lambda_i$ . The difficult task is to fulfil the exact equality  $\boldsymbol{\xi} \cdot \nabla \lambda =0$ . {In order to further simplify the notation, we define the operator $\mathcal{F}$ as the follows:
	\begin{equation}
	\mathcal{F}({\bf x}) = \boldsymbol{\xi}({\bf x}) \cdot \nabla \lambda(\bf x)
	\end{equation}
	Thus, being $\gamma$ the LCS, the condition (III) reads $\mathcal{F}(\gamma) = 0$.}
The difficulty arises from the fact that most of Lagrangian Coherent Structures lie along the ridges, where the gradients of $\lambda$ field
are very difficult to compute numerically; one should have an infinite fine grid to be able
to exactly locate the position on the ridge, and then to evaluate what is the direction of the gradient. Thus, for practical reasons the condition $\mathcal{F}(\bf x) = 0$ is replaced  by $||\mathcal{F}(\bf x) || \le \mathcal{T}$ where $ \mathcal{T}$ is an acceptance threshold.  If a point does not fulfil the condition, we explore the  neighbouring points to check whether the curve does not fulfil the condition only because of a sparse grid problem. To do that, we move a step towards the gradient of the maximum eigenvalue of the CG tensor and then we check the same condition in the new position. \\
In practice, being ${\bf x}_i$ the point of the curve that we are investigating, we analyse even the point $\hat{ {\bf x}}_i = {\bf x}_i + s\, \nabla \lambda $, {where $s$ is a "jump" parameter usually taken equal to 4. The value of $s$ will be discuss at the end of the section.} If 
the new point $\hat{ {\bf x}}_i $ has $||\mathcal{F}(\hat{ {\bf x}}_i) || \le \mathcal{T}$  or if the sign of $\mathcal{F}(\hat{ {\bf x}}_i) $ reverses with respect to $\mathcal{F}( {\bf x}_i)$ , the point ${\bf x}_i$ is considered to fulfil the LCS condition.\\
The fact that we accept even points, ${\bf x}_i$, where the sign of $\mathcal{F}( \hat{\bf x}_i({\bf x}_i))$   reverses with respect to $\mathcal{F}( {\bf x}_i)$  is due to the fact that often near ridges of $\lambda_{max}$ the gradients are not smooth and thus the changes in direction happen in infinitesimal regions. Clearly such infinitesimal regions cannot be located with a finite mesh and we detect such regions as those where the $\nabla \lambda$ reverses the component perpendicular to the tangent of vector of the LCS (namely $\xi_{min}$ by construction). \\
In the  just described procedure, we saw that there is a numerical parameter called "jump parameter". The size of this parameter says how much we move along $\nabla \lambda_{max}$ before to  compute $\mathcal{F}( \hat{\bf x}_i({\bf x}_i))$. Using too large a value could lead to a loss of physical information: false positive or false negative if the length-scales of phenomena are small. Too small a value could lead to a false negative, but not false positive. Thus, as a good practice, we use $s = 4 \, 10^{-6} N_x N_y$ (for at least $N_x N_y = 10^6$). The reader should keep in mind that even the $s$ parameter is in dimensionless units since  all the computations  are  performed on the logical grid.
To illustrate how important is condition \ref{numericallyaproblem}, let us  focus on  figure \ref{third_condd}. The picture shows three yellow curves and a vector field. The curves are obtained integrating  the equation \ref{condizione2_compu}, starting from the local maxima of $\lambda_{max}$. Thus, according to many previous works these three curves are all supposed to be LCSs. However looking at the vector field, representing the field $\nabla \lambda_{max}$, it is immediate to see that  condition (III) is not satisfied for all the curves: only one of them has its tangent vector along the gradient of the $\lambda_{max}$ field.

\begin{figure}
	\begin{tikzpicture}[trim left=-4.5cm]
	\node[node distance=1cm] (russell) at (0,0)
	{\includegraphics[height=5.5cm,width=7.5cm]{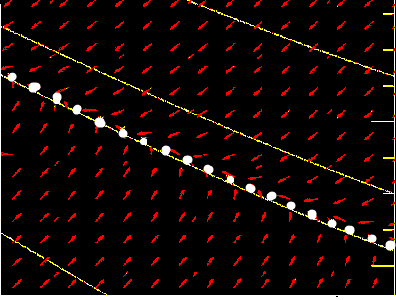}};
	\node[below of=russell, node distance=3cm, anchor=center] {{\Large{$r/a$}}};
	\node[left of=russell, node distance=4.2cm, anchor=center] {{{\Large $\theta$}}};
	\end{tikzpicture}
	\caption{three curves, in yellow, coming from condition \ref{condizione2}, the vector
		field $\nabla \lambda_{max} $ (in which each vector normalized to 1) and the points, in white, in which the
		condition 3 is satisfied (according to the algorithm described).}
	\label{third_condd}
\end{figure}

\section{Computation in cylindrical geometry} \label{App_B}
If both initial and final positions are defined in cylindrical coordinates, then  $u^i = \Theta^i$. 
In the cylindrical case ${\bf e}_r = {\bf e}^r = \hat{{\bf e}}_r, \; {\bf e}^\theta = \frac{\hat{{\bf e}}_\theta}{r}$,
$ {\bf e}_\theta = r\,\hat{{\bf e}}_\theta $.
The deformation gradient reads:
\begin{myequation} \label{deformation_gradient_cyl}
	{\bf F} = \left(
	\begin{array}{ccc}
		\frac{\partial \Phi^r }{\partial R} & \frac{1}{R} \frac{\partial \Phi ^r }{\partial \Theta} \\
		\\
		r \, \frac{\partial \Phi^\theta }{\partial R} &  \frac{r}{R} \frac{\partial \Phi^\theta }{\partial \Theta}
	\end{array}
	\right) \, .
\end{myequation}
In this tensor, $R, \, \Theta $ refer to  radial and poloidal position before deformation; $r, \, \theta $ to  radial and poloidal position after deformation. The Cauchy Green tensor is computed, in this case, with just a simple matrix product due to the fact that $\hat{g}_{i,j} = \delta_{i,j}$ ({orthogonal geometry}). For the same reason,  the change to mixed component is also the identity transformation.\\ 
Now let us  come to the integration.
In the cylindrical space, equation \ref{condizione2_compu} reads:
\begin{equation}
\frac{\text{d}r}{\text{d}s} = \xi_r \qquad \frac{\text{d}\theta}{\text{d}s} = \frac{\xi_\theta}{r} \, .
\end{equation}
Since the curve integration is performed on the logical grid,
the equation to be solved  reads
\begin{equation} \label{integration_grid}
\frac{\text{d}x_{ind}}{\text{d}s} = \hat{\xi}_r \qquad \frac{\text{d}y_{ind}}{\text{d}s} = \hat{\xi}_\theta
\end{equation}
with $\hat{\xi} = \frac{\xi_r}{\Delta r}$ and $\hat{\xi} = \frac{\xi_\theta}{\Delta \theta \, r}$, and where $\Delta r$, $\Delta \theta$ represent the discretization of the grid and $x_{ind}, y_{ind}$ the indices in the two directions.

\vspace{1cm}

\bibliography{Bibliography}

\end{document}